%% file: final_version.tex
\documentclass[fleqn,usenatbib]{mnras}

\input{additional_tex/pack}        
\input{additional_tex/definitions} 

\usepackage{newtxtext,newtxmath}

\usepackage[T1]{fontenc}

\DeclareRobustCommand{\VAN}[3]{#2}
\let\VANthebibliography\thebibliography
\def\thebibliography{\DeclareRobustCommand{\VAN}[3]{##3}\VANthebibliography}

\usepackage{graphicx}	
\usepackage{amsmath}	

\title[Little red dots and quasar clustering measurements]{``Little red dots'' cannot reside in the same dark matter halos as comparably luminous unobscured quasars}

\author[Pizzati et al.]{Elia Pizzati$^{1}$\thanks{\href{mailto:pizzati@strw.leidenuniv.nl}{pizzati@strw.leidenuniv.nl}},
Joseph F. Hennawi$^{1,2}$,
Joop Schaye$^{1}$, 
Anna-Christina Eilers$^{3}$,
Jiamu Huang$^{2}$,
\newauthor
Jan-Torge Schindler$^{4}$,
Feige Wang$^{5,6}$
\\
$^{1}$ Leiden Observatory, Leiden University, P.O. Box 9513, 2300 RA Leiden,
The Netherlands\\
$^{2}$ Department of Physics, University of California, Santa Barbara, CA 93106, USA\\
$^{3}$ MIT Kavli Institute for Astrophysics and Space Research, Massachusetts Institute of Technology, Cambridge, MA 02139, USA\\
$^{4}$ Hamburger Sternwarte, University of Hamburg, Gojenbergsweg 112, D-21029 Hamburg, Germany\\
$^{5}$ Steward Observatory, University of Arizona, 933 N Cherry Avenue, Tucson, AZ 85721, USA\\
$^{6}$ Department of Astronomy, University of Michigan, 1085 S. University Ave., Ann Arbor, MI 48109, USA\\
}

\date{Accepted XXX. Received YYY; in original form ZZZ}

\pubyear{2023}

\begin{document}
\label{firstpage}
\pagerange{\pageref{firstpage}--\pageref{lastpage}}
\maketitle

\begin{abstract}
The James Webb Space Telescope (JWST) has uncovered a new population of candidate broad-line AGN emerging in the early Universe, named ``little red dots'' (LRDs) because of their compactness and red colors at optical wavelengths. LRDs appear to be surprisingly abundant ($\approx 10^{-5}\,\cMpc^{-3}$) given that their inferred bolometric luminosities largely overlap with those of the UV-luminous quasars identified at high $z$ in wide-field spectroscopic surveys. In this work, we investigate how the population of LRDs and/or other UV-obscured AGN relates to the one of unobscured, UV-selected quasars. By comparing their number densities, we infer an extremely large and rapidly evolving obscured:unobscured ratio, ranging from $\approx20:1$ at $z\approx4$ to $\approx2300:1$ at $z\approx7$, and possibly extending out to very high ($\approx10^{47}\,\ergs$) bolometric luminosities.  This large obscured:unobscured ratio is incompatible with the UV-luminous duty cycle measured for unobscured quasars at $z\approx4-6$, suggesting that LRDs are too abundant to be hosted by the same halos as unobscured quasars. This implies that either (a) the bolometric luminosities of LRDs are strongly overestimated or (b) LRDs follow different scaling relations than those of UV-selected quasars, representing a new population of accreting SMBHs emerging in the early Universe. 
A direct comparison between the clustering of LRDs and that of faint UV-selected quasars will ultimately confirm these findings, and shed light on key properties of LRDs such as their host mass distribution and duty cycle. We provide a mock analysis for the clustering of LRDs and show that it is feasible with current and upcoming JWST surveys.
\end{abstract}




\section{Introduction}\label{sec:introduction}

The connection between the quasar phenomenon and the accretion of material onto a supermassive black hole (SMBH) was first hypothesized to account for the extraordinary luminosity inherent to quasar activity \citep[e.g.,][]{salpeter1964,zeldovich_1964,lynden_bell1969}. According to this picture, most of the accreting material contributes to growing the mass of the SMBH, but a small fraction of this material (known as the \textit{radiative efficiency}) is converted into energy and radiated away, giving rise to the quasar phenomenon. 

The argument first proposed by \citet{Soltan1982} embeds this connection into a cosmological context: integrating the total energy emitted by quasars over all cosmic time and assuming a standard radiative efficiency of $\approx 10\%$, one finds that the mass that has been accreted on black holes per unit of comoving volume up until today is  
comparable 
to the total mass density of the SMBHs we observe in the local Universe. This implies that SMBHs grew their mass while, at the same time, they were shining as active luminous quasars. 

Extensions of this argument have been employed to relate the growth of black holes to quasar activity at different cosmic times \citep[e.g.,][]{yu_tremaine2002, shankar2010_lowz}{}{}. While specific assumptions vary, these arguments are all based on the key idea that the bulk of black hole growth in the Universe is traced by the evolving demographic properties of luminous quasars. Wide-field optical spectroscopic surveys such as the Sloan Digital Sky Survey \citep[SDSS,][]{york2000} and the 2dF QSO redshift survey \citep[2QZ,][]{croom2004} examined the properties of UV-luminous, type-1 quasars, and consistently showed that quasar activity peaks around $z\approx2$ and declines rapidly towards higher redshifts \citep[e.g.,][]{richards2006, kulkarni2019}. 

UV-luminous quasars, however, are not the whole story. The radiation emitted from accreting SMBHs can be obscured by intervening dust and gas, resulting in a diverse population of Active Galactic Nuclei (AGN) whose emission properties vary greatly across the electromagnetic spectrum \citep[e.g.,][]{padovani2017}. A general dichotomy exists, however, between unobscured AGN/quasars, exhibiting a UV-optical continuum from the accretion disk, and obscured/reddened AGN whose UV emission is partly (or completely) extincted by the dust that surrounds the SMBH. 
Whether this obscuration results from a viewing-angle effect \citep{Antonucci93,UrryPadovani95} or signifies a distinct ``dust-enshrouded'' population \citep{Sanders88,Hopkins05} has been hotly debated. Nevertheless, decades of AGN censuses across the electromagnetic spectrum (optical, X-ray, mid-IR, radio) have allowed us to map the contribution of UV-obscured AGN activity as a function of redshift and AGN luminosity \citep[e.g.,][]{Ueda03, Ueda14, Merloni14,aird2015, glikman2018} 
The resulting consensus is that a significant fraction ($\approx20-80\%$)
of AGN can be obscured in the UV, even at quasar-like (intrinsic) luminosities ($L_{\rm bol} \gtrsim 10^{45}\ergs$), and that this fraction evolves mildly with redshift. Studies that include the contribution of obscured AGN environments to the total SMBH growth budget \citep[e.g.,][]{hopkins2007,shen2020}
support the general picture outlined by the Soltan argument, pointing to a radiative efficiency for accretion on SMBHs close to $\approx 10\%$, and indicating that the bulk of SMBH growth took place during cosmic noon ($z \approx 1-3$).  
 
While a multi-wavelength exploration of AGN activity is possible at $z\lesssim3$, our understanding of black hole growth and accretion in the high-redshift Universe ($z\gtrsim4$) has been informed almost exclusively by the population of UV-luminous, type-1 quasars detected by optical/NIR wide-field surveys up to $z\approx7.5$ \citep[e.g.,][]{fan2022}. This population is commonly assumed to trace the underlying evolution of AGN/SMBH activity (including UV-obscured sources) at high $z$ by simply extrapolating the obscuration properties of quasars from low/intermediate redshifts \citep[e.g.,][]{shen2020}. Whether this extrapolation is reliable and can offer an unbiased view of SMBH growth and AGN activity in the first billion years of the Universe is currently unclear.
Several simulations  \citep[e.g.,][]{ Ni2020, Vito2022, Bennett2024} and observations 
\citep{Vito18,Circosta19,DAmato20,Gilli2022}, for example, have suggested a rapid evolution of the obscuration properties of quasars/AGN in the early Universe, due to the presence of high-column density gas within the innermost regions of their host galaxies.

The advent of the James Webb Space Telescope (JWST) marks a huge step forward in the study of AGN activity and SMBH growth in the early Universe. JWST has the sensitivity to go beyond the UV-selected quasar population that has been studied for decades \citep[e.g.,][]{fan2022}. Indeed, early results are already causing a seismic shift in our understanding of AGN populations at high $z$: photometric and spectroscopic JWST surveys are uncovering surprisingly large samples of faint AGN candidates at $z\approx4-10$ \citep[e.g.,][]{harikane2023,Maiolino23,maiolino2024,ubler2023,kocevski2023,kokorev2023,scholtz2023,matthee2024,greene2024, bogdan2024,kocevski2024,mazzolari2024, furtak2024, taylor2024}.
Although selection methods vary, the most reliable candidates are identified via broad H$\alpha$ or H$\beta$ lines. These lines can be used to infer AGN luminosities of $L_\mathrm{bol}\gtrsim 10^{44-45}\,\ergs$ and black hole masses of $M_\mathrm{BH}\gtrsim10^{6-7}\,\msun$ 
These masses and luminosities vastly extend the range of AGN properties that we can probe at high $z$, offering key insights on the co-evolution of SMBHs and their host galaxies \citep[e.g.,][]{inayoshi2022,pacucci2023_mbhmstar}{}{}, the contribution of AGN to hydrogen reionization \citep[e.g.,][]{Maiolino23, dayal2024, madau2024}, and potentially also on SMBH seeding/growth models \citep[e.g.,][]{pacucci2022, li2024_bh_growth}. 

Yet, relating this new population of JWST AGN to the one of UV-selected high-$z$ quasars has proven challenging. Even though they generally resemble standard, type-1 quasars at rest-frame optical wavelengths, JWST broad-line AGN appear to be much more abundant than what was expected by extrapolating the quasar luminosity function (QLF) to faint UV luminosities \citep[][]{harikane2023}. It is currently unclear whether QLF studies have been strongly underestimating the number of faint UV quasars that are present at high $z$ \citep[e.g.,][]{giallongo2019}, or whether the AGN population revealed by JWST using broad optical lines presents substantially different properties from those of UV-selected, type-1 quasars, as also suggested by their peculiar Spectral Energy Distribution (SED) features such as X-ray weakness \citep[][]{maiolino2024, lambrides2024} and (tentative) lack of variability \citep[][]{mitsuru2024}. Upcoming JWST surveys will probe the properties of these broad-line AGN in the rest-frame UV, providing key insight into their nature and allowing a direct comparison to the UV-selected quasar population.

Interestingly, however, some of the AGN revealed by JWST are even more remarkable: a significant fraction of them ($\gtrsim20\%$; \citealt{harikane2023,taylor2024}) show a 
steep red continuum in the rest-frame optical pointing to moderate dust reddening values of 
$A_\mathrm{V}\approx1-4$ \citep[][]{kokorev2024, greene2024}{}{}. When correcting for the attenuation of dust to the continuum and/or broad-line emission, these obscured/reddened AGN have inferred bolometric luminosities of $L_{\rm bol} \approx 10^{45-46}\,\ergs$ and SMBH masses up to $\approx10^{7-8}\,\msun$ \citep[][]{greene2024, kocevski2024, harikane2023}. Hence, they largely overlap in luminosity and SMBH mass with the population of UV-selected, type-1 quasars revealed in pre-JWST surveys \citep[][]{fan2022, matsuoka2022}. 
This is incredibly surprising, since these UV-luminous quasars with comparable luminosities (and redshifts) were selected from wide-field 1400 deg$^2$ deep 
imaging surveys probing volumes of $\approx 10^{10}~{\rm cMpc}^3$ \citep{matsuoka2022}, whereas JWST AGN 
are identified in surveys of not more than $\approx 300-600~{\rm arcmin}^2$ probing a volume not
greater than $\approx 10^6-10^7~{\rm cMpc}^3$ \citep[][]{matthee2024,kokorev2024}. Such a massive difference indicates 
that these AGN may be tracing a new population of broad-line, obscured sources\footnote{Standard AGN classifications \citep[e.g.,][]{padovani2017} divide low-$z$ quasars in type-1 (showing broad emission lines in their spectra) and type-2 (showing only narrow emission lines). Type-2 quasars are generally identified with obscured sources whose broad lines are extincted by dust. Even though their continuum is heavily reddened at optical and UV wavelengths, JWST AGN are always revealed by broad optical lines, and hence they officially belong to the type-1 quasar category. While examples of type-1, reddened quasars exist at low redshifts, they are rare compared to the global quasar population (Wang et al., in prep.), making the interpretation of these new JWST AGN sources even more challenging.} that are far more abundant than comparably luminous UV-unobscured quasars. According to this picture, our understanding of SMBH growth and quasar/AGN activity at high-$z$ -- which was entirely based on the demographic properties of UV-luminous quasars -- needs to be thoroughly revised to account for this new, large AGN population that 
is in place in the early Universe \citep[e.g.,][]{inayoshi2024, li2024}. 

 As shown by \citet{greene2024}, the reddened broad-line AGN in JWST surveys tend to have a 
characteristic v-shaped 
 SED, with the red continuum in the rest-frame optical transitioning to relatively blue colors in the rest-frame UV. 
While the physical origin of this SED shape is currently unclear 
\citep[e.g.,][]{killi2024,li2024, wang2024, kokorev2024b, inayoshi_maiolino2024},
several studies have exploited these peculiar SED features and applied specific color and compactness cuts to NIRCam photometry to isolate obscured broad-line AGN photometrically \citep[e.g.,][]{labbe2023, perez-gonzalez2024,kokorev2024,kocevski2024,akins2024}. By applying similar photometric selections, \citet{greene2024} and \citet{kocevski2024} have proved that a large fraction 
of the selected sources ($\gtrsim70-80\%$) is indeed comprised of reddened, high-redshift ($z\approx4-8$), broad-line AGN. 
Sources selected using these methods have become known as ``Little Red Dots'' (LRDs henceforth; \citealt{matthee2024}) because of their compactness and peculiar colors in NIRCam imaging. We note that this term has been used in the literature to refer to samples obtained following different spectroscopic and photometric criteria. 
Here, with the term ``Little Red Dots'' we refer to the above-mentioned population of candidate broad-line AGN that are red at optical wavelengths, and hence have quasar-like inferred bolometric luminosities and black hole masses. We include in our analysis both spectroscopic \citep[][]{greene2024} and photometric \citep[][]{kokorev2024} samples: while the latter may be subject to a significant degree of contamination \citep[e.g.,][]{taylor2024}, their number densities agree well with the ones from spectroscopy \citep[][]{greene2024}\footnote{On top of that, a moderate degree of contamination does not impact the main conclusions of our analysis (see Sec. \ref{sec:conclusions} for further discussion).}. We mention the caveat, however, that even for spectroscopically confirmed broad-line LRDs, the presence of an accreting SMBH and the nature of the observed SED are still heavily debated \citep[e.g.,][]{durodola2024, li2024, perez-gonzalez2024, ananna2024, yue2024, maiolino2024, mitsuru2024, baggen2024, inayoshi_maiolino2024}. In the following, we \textit{assume} that LRDs are obscured, broad-line AGN, and examine the consequences of the large obscured:unobscured ratio at quasar-like bolometric luminosities that is implied by this assumption. We defer the reader to Sec. \ref{sec:conclusions} for a discussion on the nature of LRDs and the conclusions we can draw from our results. There, we will also examine how the general population of faint (unobscured) broad-line AGN revealed by JWST \citep[e.g.,][]{harikane2023, maiolino2024, taylor2024} fits in the discussion presented in this work. 

If a huge obscured LRD population is indeed present at high redshifts, 
the first question that awaits to be answered is: how does this population compare to that of comparably luminous, UV-selected quasars in terms of SMBH mass and accretion rate, host 
environments, and evolution history? Are LRDs standard, actively accreting quasars whose emission is attenuated by intervening dust and gas, or do they represent a different evolutionary stage in the accretion history of SMBHs? Are UV-luminous quasars and LRDs drawn from the same population of halos/galaxies?

In this work, we take a first step towards answering these questions by studying the properties of quasars and LRDs in terms of their number density and large-scale environment/host halo mass. In particular, we argue that the extreme abundance of LRDs/obscured AGN is at odds with the duty cycle of UV-luminous 
quasar activity at $z\approx4-6$ inferred from the combination of quasar clustering and luminosity function measurements
\citep[][]{shen2007,eilers2024,pizzati2024a, pizzati2024b}. 
This indicates that LRDs cannot be drawn from the same population of dark matter halos as UV-selected quasars, notwithstanding that quasars and LRDs have the same inferred bolometric luminosities and SMBH masses. Hence, provided that these luminosities and masses are indeed correct, LRDs would need to obey fundamentally different scaling relations than the ones holding for quasars, as the same SMBH masses are linked to smaller host halo masses. Possibly, this points to the fact that LRDs represent a different evolutionary stage in the accretion history of SMBHs at early cosmic time. 

In order to support these conclusions and unveil the accretion history and large-scale environment of LRDs, measuring the clustering of these sources is key. Here, we suggest that a convincing measurement of the duty cycle and host halo mass of LRDs can be obtained by using NIRCam/WFSS observations of LRD fields and measuring the cross-correlation between LRDs and \OIII line emitters, with a similar setup and strategy to current JWST programs targeting UV-luminous, high-$z$ quasars, such as EIGER \citep[][]{kashino2022, eilers2024} and ASPIRE \citep[][]{aspire_wang2023}. Using the methodology developed in previous work \citep[][]{pizzati2024a,pizzati2024b}, we provide a mock analysis for these clustering measurements and discuss the prospect of undertaking this measurement with current and future JWST programs.

The paper is structured as follows. In Sec. \ref{sec:lrd_abundance}, we compare the abundance of LRDs/obscured AGN with the one of the 
UV-luminous high-$z$ quasar population, inferring a large and rapidly evolving obscured:unobscured ratio at $z\approx4-8$. 
Sec. \ref{sec:clustering_duty_cycle} studies the implications of this large ratio in terms of host dark matter halo populations, 
and points to clustering studies as a way to determine the nature of LRDs. Sec. \ref{sec:lrd_mocks} 
provides a mock analysis of this clustering measurement. The results are discussed and summarized in Sec. \ref{sec:conclusions}.

\section{The Staggeringly High Abundance of UV-Obscured AGN Implied by little red dots} \label{sec:lrd_abundance}

\begin{figure*}
	\centering
	\includegraphics[width=\textwidth]{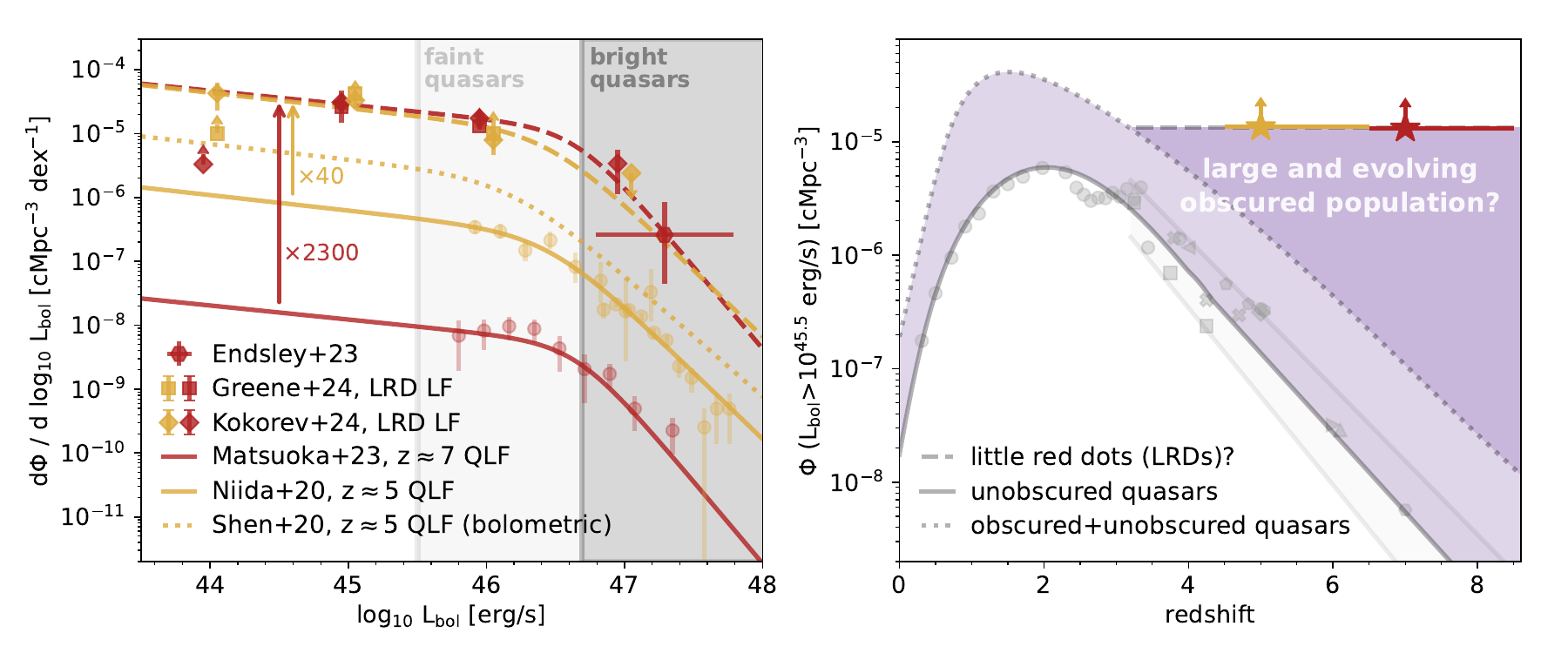}
	 \caption{ 
  \textit{Left}: Luminosity function of UV-selected quasars, expressed in terms of bolometric luminosities, compared to the bolometric luminosity function of Little Red Dots (LRDs) at different redshifts. Solid lines show the fits to the unobscured quasar luminosity functions (QLFs) at $z\approx5$ (\citealt{niida2020}; golden color) and $z\approx7$ (\citealt{matsuoka2023}; red). Data points for these QLFs are also shown as circles. The bolometric QLF compiled by \citet{shen2020} at $z\approx5$ is shown with a dotted line. Bolometric luminosity functions (LFs) for LRDs are shown with square \citep[][]{greene2024} and diamond \citep[][]{kokorev2024} symbols. Golden (red) symbols refer in this case to the redshift range $4.5<z<6.5$ ($6.5<z<8.5$). The number density implied by the single source identified by \citet{Endsley22,Endsley23} at $z\approx7$ (see main text) is shown as a red hexagon. Vertical arrows show by how much the QLF fits (solid lines) need to be rescaled to match the LRD luminosity functions. Dashed lines show the rescaled QLFs: the $z\approx5$ ($z\approx7$) QLF is rescaled by a factor of $40$ ($2300$). The light (dark) grey shaded region highlights the luminosity range $10^{45.5}\,\ergs<L_\mathrm{bol}<10^{46.5}\,\ergs$ ($L_\mathrm{bol}>10^{46.5}\,\ergs$).
  \textit{Right:} Evolution of the number density of quasar/AGN (above the luminosity threshold $L_\mathrm{bol}>10^{45.5}\,\ergs$) with redshift. Gray points show the number densities obtained by integrating individual fits to the unobscured QLFs above the luminosity threshold (partly adapted from the compilation in \citealt{schindler2023}; fits are taken from \citealt{yang2016,akiyama2018,mcgreer2018,matsuoka2018,schindler2019,kulkarni2019, niida2020, onken2021, pan2022, schindler2023, matsuoka2023}). The solid line shows an evolutionary model for the unobscured quasar number density obtained by smoothly interpolating between the fit of \citet{kulkarni2019} at $z<4$, and an exponential decline $\Phi\propto10^{-kz}$, with $k=0.7$, at higher redshifts. The gray shaded area and the two lines at $z>3$ are meant to bracket our uncertainty on the number density of high-$z$ unobscured quasars.  Dotted lines show the number density evolution predicted by the bolometric luminosity function of \citet{shen2020} (see their ``global fit B''). Colored star symbols show the number density for the LRDs obtained by integrating the rescaled QLFs from the left panel. The flat evolution of the LRD number density implied by the data points is highlighted with a horizontal dashed line, while the (light+dark) purple-shaded areas show the AGN obscured:unobscured ratio inferred from LRDs and low-$z$ multi-wavelength observations.
  \label{fig:overview}
 	}
\end{figure*}

In this section, we compare the luminosity function 
of the UV-luminous, unobscured population of quasars to that of the new population of UV-obscured ``Little Red Dots'' (LRDs) uncovered in JWST surveys. Our goal is to study the abundance of these two populations across cosmic time, and infer an estimate of the AGN obscured fraction at different redshifts. 

To this end, we use bolometric luminosities as a way to probe the intrinsic radiation emitted by the different quasar/AGN populations 
prior to any obscuration effects.  The bolometric luminosities of UV-luminous, type-1 quasars can 
be easily inferred from their UV-continuum absolute magnitude by assuming standard bolometric correction factors that are available in the 
literature \citep[e.g.,][]{richards2006, runnoe2012, shen2020}. In this work, we use the relation between the $M_{1450}$ absolute 
magnitude and the bolometric luminosity $L_\mathrm{bol}$ presented in \citet{runnoe2012}\footnote{The bolometric correction for 
$\lambda=1450$ \r{A} 
is $\log_{10} L_{\rm iso}/\ergs  = 4.745 + 0.910 \log_{10} \lambda L_\lambda/\ergs$. $L_{\rm iso}$ refers to the bolometric 
luminosity computed under the assumption of isotropy, and it is related to the observed bolometric luminosity 
$L_\mathrm{bol}$ through the relation $L = 0.75 \, L_{\rm iso}.$}. While other bolometric correction factors may return 
slightly different results because of the choices made for the quasar SED and the parametrization of the UV-bolometric relation, 
the uncertainty in the bolometric correction for UV-selected, type-1 quasars is relatively small and has little impact on our conclusions.

Estimating the intrinsic bolometric luminosity of the LRD population, instead, is much more challenging. While bolometric luminosities are easy to constrain for UV-selected quasars because one directly probes the big-blue-bump (where the bulk of the emission comes out, \citealt{sanders1989}), dust obscuration prevents a direct determination of the LRD luminosities from their UV emission. For low-$z$, dust-obscured quasars, it is usually possible to constrain the radiation reprocessed by dust in the mid-IR with Spitzer \citep[e.g.,][]{lacy2015}. However, this is currently not a viable option for LRDs, as they appear to manifest only at high $z$ and the bulk of their expected mid-IR emission is redshifted to wavelengths of $\approx70\mu\mathrm{m}$, which are not accessible from the ground and are only probed by shallow surveys (e.g., Herschel). The only option that remains available for estimating the bolometric luminosities of LRDs is to use the emission in the optical continuum and/or broad optical lines and convert that to a bolometric luminosity using some scaling relations \citep[e.g.,][]{richards2006, runnoe2012b}, which are however fairly uncertain. Even more relevantly, one has to properly account for the effects of dust obscuration on the observed optical emission. Current estimates of the bolometric luminosities 
for the LRD population \citep[e.g.,][]{greene2024, kokorev2024, akins2024} 
rely on the assumption that the optical continuum of LRDs is dominated by dust-reddened AGN radiation and use the slope of the SED in the optical continuum to infer the amount of obscuration in place. However, this continuum emission could be contaminated by radiation from the host galaxy: disentangling the contributions of the central SMBH and the stellar light to the SED of LRDs is currently a hotly debated problem \citep[e.g.,][]{ durodola2024, li2024, perez-gonzalez2024,baggen2024, inayoshi_maiolino2024}. 
As mentioned before, here we simply assume that bolometric luminosity 
estimates for LRDs are correct.
A discussion on how our results are impacted by uncertainties in the bolometric 
luminosities of LRDs can be found in Sec. \ref{sec:conclusions}.

In the left panel of Figure \ref{fig:overview}, we show the luminosity function of UV-luminous, unobscured quasars (expressed in terms of 
bolometric luminosities) at two sample redshifts of  $z\approx5$ (\citealt{niida2020}; golden solid line and points) 
and 
$z\approx7$ (\citealt{matsuoka2023}; red solid line and points). These luminosity functions can be compared to the bolometric 
luminosity functions of LRDs measured by \citet{greene2024} (squares) and \citet{kokorev2024} (diamonds)\footnote{
The \citet{greene2024} luminosity function is obtained from a small sample of spectroscopically-confirmed broad-line LRDs in the UNCOVER field \citep[][]{bezanson2022}. The work of \citet{kokorev2024} applies the photometric selection suggested by \citet{labbe2023} and \citet{greene2024} to a larger sample of JWST blank fields, identifying 260 AGN candidates in $\approx640
$ arcmin$^2$ of JWST imaging. 
While several other LRD luminosity functions have been published in the literature \citep[see e.g.,][]{matthee2024, kocevski2024, lin_aspire2024}, 
none of these are based on \textit{unattenuated} bolometric luminosities. Accounting for the effect of dust attenuation is key if our goal is to compare the luminosities of LRDs to the ones of UV-luminous quasars. The only exception is the recent work of \citet{akins2024}, who also published an LRD bolometric luminosity function corrected for obscuration effects. However, their photometric selection differs significantly from the one presented in \citet{greene2024} and \citet{kokorev2024}, and hence we do not include their sample in the analysis. We note however that they find even larger number densities for LRDs, which would strengthen our conclusion on the presence of a large obscured high-$z$ AGN population.}. Golden (red) symbols refer to the redshift range $4.5<z<6.5$ ($6.5<z<8.5$).
This plot highlights the strikingly different abundance of LRDs compared to the UV-luminous quasar population. As also mentioned in the introduction, 
this difference reflects the fact that LRDs are common in the small fields ($\approx300-600$ arcmin$^2$) probed by JWST surveys, whereas unobscured quasars are notoriously rare and can be sampled only by wide-field surveys of $\approx2000$ deg$^2$. 

By directly comparing the luminosity functions of UV-luminous quasars and LRDs, we can quantify the different abundances of these two populations as a function of their luminosity. Interestingly, we find that the \textit{shape} of the LRD luminosity function resembles the one of the UV-luminous quasar luminosity function (QLF) at both redshifts. Indeed, if we scale up the \citet{niida2020} fit to the $z\approx5$ QLF by a factor of $\approx40$, we get a good match to the LRD luminosity function in the redshift range $4.5<z<6.5$. This suggests that LRDs may constitute a new, obscured population of accreting SMBHs at $z\approx5$, outnumbering unobscured quasars by $\approx$40:1 at all luminosities. Similar -- but even more extreme -- conclusions can be drawn at $z\approx7$. In this case, the fit to the \citet{matsuoka2023} QLF needs to be scaled up by a factor of $\approx2300$ to match the LRD luminosity function at $6.5<z<8.5$, implying an even larger obscured:unobscured ratio, 
roughly independent of luminosity. 

We note that care must be taken to extend these conclusions to a large range of bolometric luminosities. Most LRDs have inferred (dust-corrected) bolometric 
luminosities in the range $\approx10^{44-46}\,\ergs$. The faintest high-$z$ unobscured quasars identified in wide field surveys have luminosities of $\approx10^{45.3}\,\ergs$ \citep[e.g.,][]{matsuoka2022}. 
Hence, a proper 
comparison between LRD and quasar number densities can be carried out only for the \textit{bright} population of 
LRDs with $L_\mathrm{bol}\approx10^{45.5-46.5}\,\ergs$. At lower bolometric luminosities, the UV-luminous QLFs are only based on 
extrapolations; hence, conclusions on the obscured fraction of faint ($L_\mathrm{bol}\lesssim10^{45} \,\ergs$) AGN are only tentative. 
At very bright luminosities of $L_\mathrm{bol}\approx10^{47}\,\ergs$, the number density of UV-luminous quasars is very well 
constrained \citep[e.g.,][]{schindler2023}. Very bright LRDs, on the other hand, are hard to find in the small field of views 
(FoVs) probed by JWST surveys and the only constraints we have on their number density come from the work of \citet{kokorev2024} 
(see also \citealt{akins2024}), which is however only based on photometry with no spectroscopic confirmation.

Interestingly, signs of a large obscured AGN population at high bolometric luminosities ($L_{\rm bol} \gtrsim 10^{47}\,\ergs$) come from different data.  
Using multi-wavelength observations in mid-/far-IR, sub-mm, and radio, \citet{Endsley22,Endsley23} (see also \citealt{Lambrides23})
discovered an extremely luminous ($L_{\rm bol}=(2.0\pm0.2)\times10^{47}\,\ergs$) obscured, radio-loud quasar at $z=6.83$ 
in just $1.5\,{\rm deg}^2$ of COSMOS imaging, and argued for an extremely large obscured:unobscured ratio of $\sim 2000:1$.
We can get an estimate of the number density implied by this source by simply computing the total 
comoving volume in the COSMOS field for the redshift range $6.6<z<6.9$ (in which the source was photometrically selected; 
see \citealt{Endsley22}). We get a volume of $3.8\times10^6\,\cMpc^3$ and a number density of $2.6\times10^{-7}\,\cMpc^{-3}$. 
For reference, we add this source to the luminosity function plot of Fig. \ref{fig:overview} (left), by assuming 
a 1 dex bin in bolometric luminosity centered on the quasar's measured $L_\mathrm{bol}$. Upper and lower limits are computed 
assuming Poisson statistics for a single source \citep[see][]{gehrels86}. Despite the large uncertainties, this source supports the existence of a large obscured population at the bright end of the QLF compatible with the one found for LRDs. 

In what follows, we will consider two separate hypotheses: (a) there is a large obscured AGN/quasar population at bolometric 
luminosities $L_\mathrm{bol}\approx10^{45}-10^{46}\,\ergs$ (i.e., at the faint end of the quasar luminosity function; 
light-grey shaded area in the left panel of Fig. \ref{fig:overview}); (b) this large obscured population extends to 
very large bolometric luminosities of $L_\mathrm{bol}\approx10^{47}\,\ergs$ (dark-grey shaded area). While the former is 
supported by a fairly large sample of LRDs that have been argued to overlap in luminosity with the faint quasar population 
(e.g., \citealt{greene2024, matthee2024, lin_aspire2024, taylor2024}; Schindler et al, in prep.), the latter is currently based only on a handful of sources (i.e., the photometrically-selected LRDs in \citealt{kokorev2024, akins2024} and the obscured quasars from \citealt{Endsley22,Endsley23,Lambrides23})
and thus it is only tentative 
(see Sec. \ref{sec:conclusions} for further discussion).  

In the right panel of Figure \ref{fig:overview}, we show how the quasar/AGN number density evolves with redshift by integrating the QLF above a bolometric luminosity threshold of $L_\mathrm{bol} = 10^{45.5}\,\ergs$ (light grey vertical line in the left panel). The cosmological evolution of the UV-luminous, type-1 quasar population has been analyzed in the recent work of \citet{kulkarni2019}. The solid grey line in Fig. \ref{fig:overview} (right) shows their best-fitting model at $z<4$. For higher redshifts, the \citet{kulkarni2019} model is very uncertain and does not agree well with the data. For this reason, at $z>4$ we assume that the cosmic number density of unobscured high-$z$ quasars declines exponentially as $\Phi(z)\propto10^{-kz}$, and set $k=0.7$ for our fiducial model \citep[][]{schindler2023}. We then smoothly interpolate between the fit of \citet{kulkarni2019} at $z<4$ and this exponential decrease at higher redshift. Together with this global evolution model, we also show individual (gray) points obtained by integrating local fits to the QLFs above the luminosity threshold (see the legend for references). 
Overall, these individual data points agree with the global evolutionary model, but a significant spread is present due to uncertainties in the QLF measurements 
(especially at the faint end, $L_\mathrm{bol}\lesssim10^{46}\,\ergs$). 
To quantify this uncertainty, we plot two gray lines corresponding to different exponential declines of the quasar number density, 
$k=0.65$ and $k=0.78$ \citep[e.g.,][]{wang2019, matsuoka2023}; 
these two lines are normalized at $z=4$ to twice and half of the fiducial 
model, respectively.

Together with the measurements for the UV-luminous quasar number density, we show (Fig. \ref{fig:overview}, right panel) with a dotted line the model for the evolution of the AGN \textit{bolometric} number density from \citet{shen2020}. This work employs multi-wavelength observations (from X-rays to mid-IR) to include the contribution of all quasars/AGN to the number density budget. In particular, by exploiting X-ray 
observations at $0<z<3$ \citep[e.g.,][]{Ueda03, Ueda14, Merloni14, aird2015}, they include a model for AGN obscuration, and account 
for the obscured fraction of quasars/AGN in their luminosity function estimates. As mentioned in the introduction, observations generally constrain the AGN obscured fraction only at $z\lesssim3$, so the \citet{shen2020} model is effectively extrapolating the behaviour of the AGN obscured populations from cosmic noon to the high $z$ Universe. Nonetheless, the work of 
\citet{shen2020} represents our best guess (prior to JWST observations) for how the global AGN/SMBH population evolves as a function of redshift.
By comparing the number density of UV-selected quasars (solid grey line in the right panel of Fig. \ref{fig:overview}) with the 
number density from \citet{shen2020} (which includes obscured sources), we can estimate the obscured:unobscured ratio of AGN as a 
function of redshift. The same ratio can be studied as a function of intrinsic luminosity by considering the UV-luminous and the bolometric 
QLFs at a single redshift. As an example, we do this in the left panel of Fig. \ref{fig:overview} by showing the \citet{shen2020} 
predictions for the bolometric QLF at $z=5$ with a golden dotted line. In general, the obscured:unobscured ratio 
implied by comparing the bolometric \citep{shen2020} to the UV \citep[][]{kulkarni2019} QLFs
evolves 
moderately with redshift and luminosity, ranging from $\approx$ a few$:1$ up to $\approx 20:1$ for the case of high redshift and low bolometric luminosity. 
We note that these values are inevitably 
very uncertain, as the method employed here is subject to the exact 
parametrizations employed by \citet{kulkarni2019} and 
\citet{shen2020} for their respective QLFs. Nevertheless, we present this 
comparison 
between UV-selected and bolometric models 
to outline the conventional wisdom on AGN/quasar populations that is being challenged by the new population of LRDs/broad line AGN uncovered in JWST surveys.

The number density evolution of LRDs can be estimated by integrating their bolometric luminosity functions in the left panel of Fig. \ref{fig:overview} above the same $L_\mathrm{bol}$ threshold of $10^{45.5}\,\ergs$ employed before (vertical light grey line). In practice, given that the rescaled UV QLFs (dashed lines in the left panel of Fig. \ref{fig:overview}) are good fits to the LRD bolometric luminosity functions, we can simply rescale the unobscured quasar number density obtained at $z=5$ and $z=7$ to get the LRD number densities at the same redshifts. We show as colored star symbols (Fig. \ref{fig:overview}, right panel) the LRD number densities obtained after this rescaling. Following \citet{greene2024}, we plot these symbols as lower limits.

As argued before, the AGN number density implied by JWST observations of LRDs is surprisingly large and non-evolving. To highlight this behavior, we plot (Fig. \ref{fig:overview}, right panel) a horizontal dashed line for $z\gtrsim3$ corresponding to the abundance $\Phi_\mathrm{LRD}\approx1.3\times10^{-5}\,\cMpc^{-3}$. At $z\gtrsim6$, this abundance is many orders of magnitude higher than the one measured for unobscured quasars, implying that our general understanding of SMBH accretion and quasar activity in the early Universe may need to be deeply revised. 
\citet{inayoshi2024} 
(see also \citealt{akins2024}) 
have already examined the challenges that these LRD number densities pose to our paradigm of SMBH growth as well as the co-evolution of SMBHs and galaxies. In this work, we focus on the consequences of the large and rapidly evolving
 AGN obscured fraction that can be inferred by comparing LRDs to unobscured quasars. In Fig. \ref{fig:overview} (right), 
 we show with a light purple shading the region between the unobscured quasar evolution model and the bolometric 
 (obscured+unobscured) model of \citet{shen2020}. A darker shading highlights the dramatic increase in the obscured 
 fraction at $z\gtrsim4$ that is needed to match LRD measurements. 

Dividing the LRD number density, $\Phi_\mathrm{LRD}$ (which, to a first approximation, is not evolving with redshift), by the number density of UV-luminous quasars (solid grey line in the left panel of Fig. \ref{fig:overview}), we infer an obscured:unobscured ratio that increases from $r_\mathrm{obsc}\approx20^{+20}_{-10}:1$ at $z=4$ to 
$r_\mathrm{obsc}\approx2300^{+3500}_{-1400}:1$ at $z=7$. In the following section, we will also make use of 
the obscured:unobscured ratio at $z=6.25$, which is $r_\mathrm{obsc}\approx815^{+1600}_{-545}:1$. 
The uncertainties on these obscured ratios are computed by considering the grey shaded area 
(and grey lines) in Fig. \ref{fig:overview} (right), and are meant to quantify the scatter 
(coming from systematics in the QLF modeling) between different number density measurements for the 
unobscured quasar population. Given the challenges with interpreting and contextualizing LRD measurements, 
we currently do not attempt to model uncertainties for the LRD population, and defer to Sec. \ref{sec:conclusions} for a discussion of the significance of our results.

\section{Little red dots and UV-selected quasars: do they belong to the same population?} \label{sec:clustering_duty_cycle}

\begin{figure*}
	\centering
	\includegraphics[width=0.95\textwidth]{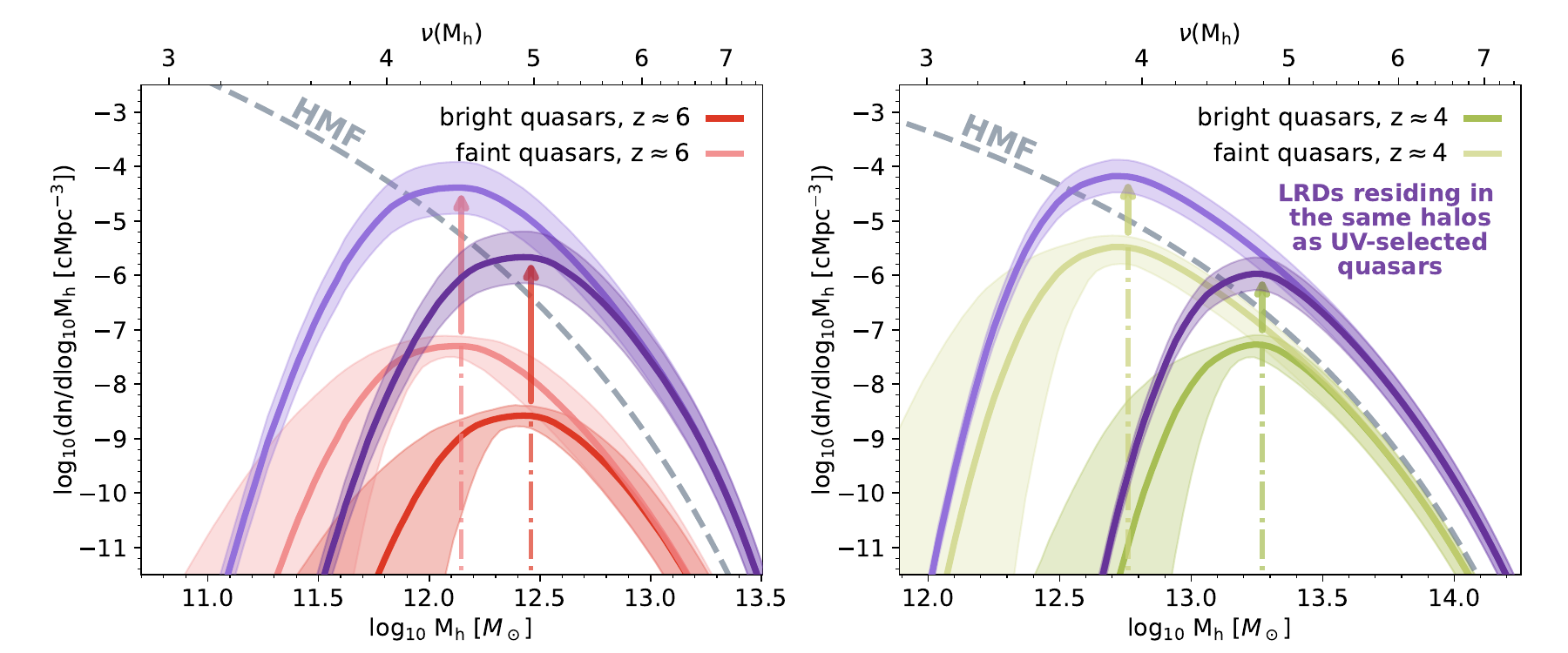}
	 \caption{Quasar-host mass functions (QHMFs) at $z\approx6$ (\textit{left panel}) and $z\approx4$ (\textit{right}) as a function of the (sub)halo mass, $M_\mathrm{h}$, and the peak height, $\nu(M_\mathrm{h})$ (see text for definitions). Darker (lighter) colors represent QMHFs obtained by setting a bolometric luminosity threshold corresponding to that of bright (faint) quasars, i.e., $L_\mathrm{bol}=10^{46.7}\,\ergs$ ($L_\mathrm{bol}=10^{45.5}\,\ergs$). Median and $1\sigma$ uncertainties (obtained by randomly sampling the posterior distributions shown in \citealt{pizzati2024a, pizzati2024b}), are represented with solid lines and shaded areas, respectively.  The dashed-dotted lines highlight the median values of the QHMF distributions.
     The (sub)halo mass functions (HMFs) at the respective redshifts are plotted with dashed grey lines in both panels. Purple colors show the QHMFs distributions when scaled up by the obscured:unobscured ratios ($r_\mathrm{obsc}$) derived in Sec. \ref{sec:lrd_abundance}, and represent the host mass distribution of LRDs under the hypothesis that they are drawn from the same halo population as UV-selected quasars.  Dark (light) purple is associated with bright (faint) quasar bolometric luminosities. The colored arrows represent the values of $r_\mathrm{obsc}$ by which the QHMFs are scaled up. The (purple) shaded regions represent the effect of the uncertainties on $r_\mathrm{obsc}$ (see Sec. \ref{sec:lrd_abundance}). The host mass distributions for LRDs overshoot the HMFs at the massive end, implying that LRDs are too abundant to reside in the same dark matter halos as comparably luminous, unobscured quasars. 
  \label{fig:qhmfs}
 	}
\end{figure*}

From the analysis performed in the previous section, we concluded that: (a) LRDs imply the existence of a large and rapidly evolving obscured 
AGN population (at redshifts $z\approx 4-7$ and bolometric luminosities $L_\mathrm{bol}\approx10^{45}-10^{46}\,\ergs$) which outnumbers 
UV-luminous quasars by several orders of magnitude (Fig. \ref{fig:overview}, right); (b) there is tentative evidence 
(Fig. \ref{fig:overview}, left) that this obscured population extends to even higher bolometric luminosities 
($L_\mathrm{bol}\approx10^{47}\,\ergs$). In this section, we examine the implications of these findings in the context of AGN host dark matter halo masses and duty cycles. 

\subsection{The host dark matter halos and duty cycles of high-$z$ unobscured quasars and their luminosity dependence} \label{sec:unobscured_quasars}

Determining which halos can host quasar activity 
as a function of cosmic time is one of the main questions in the field, as it is key to embedding quasars in the structure formation picture: this sheds light on 
the processes governing SMBH accretion and growth, as well as the co-evolution between SMBHs 
and their host halos/galaxies. In this context, quasar clustering measurements have been widely 
used to estimate the masses of the halos hosting UV-luminous quasars at different redshifts 
\citep[][]{porciani_2004,croom2005,porciani_norberg2006,shen2007,ross2009,eftekharzadeh2015, arita2023, eilers2024}{}{}. The 
idea behind these measurements is straightforward: according to the ${\rm \Lambda}$CDM cosmology, the clustering of any populations of 
objects increases with the masses of the dark matter halos they reside in \citep[e.g.][]{kaiser1984, bardeen1986, mo_white1996}.  

As pointed out by, e.g., \citet{martini2001, haiman_hui2001}, 
determining the quasars' characteristic host halo masses can also give us insight into their accretion history. 
Suppose that -- as routinely assumed -- all massive halos host a SMBH at their center. The \textit{duty cycle} of quasar activity 
determines what fraction of these SMBHs, on average, are active as UV-luminous quasars at any given moment. By comparing the number density of 
potential
quasar hosts -- obtained from quasar clustering measurements -- to the observed unobscured quasar number density, 
one can constrain this \textit{UV-luminous quasar duty cycle}. Given the connection between quasar activity and SMBH accretion and growth, the 
quasar duty cycle offers a direct view into the growth mode of SMBHs at a given cosmic epoch.

In \citet{pizzati2024a, pizzati2024b}, we 
developed a method to constrain the UV-luminous quasar duty cycle ($\varepsilon_\mathrm{QSO}$) as well as the mass distribution of the (sub)halos that host unobscured quasars (the so-called ``\textit{quasar host mass function}''; QHMF) by simultaneously fitting the clustering of quasars and their luminosity function. The method builds on a conditional luminosity function (CLF) framework, which links in a statistical sense the population of dark matter subhalos to that of quasars \citep[e.g.,][]{yang2003, ren_trenti2020}. We employ a description for the CLF based on an empirical relation between the quasar bolometric luminosity, $L_\mathrm{bol}$, and the host (sub)halo mass, $M_\mathrm{h}$, with log-normal scatter, $\sigma$. This relation is also normalized by an active fraction, $f_\mathrm{on,UV}$, which accounts for the fact that not all quasars are actively accreting and UV-luminous at a given time:
\begin{equation}
\begin{split}
        {\rm CLF}&(L_\mathrm{bol}| M_\mathrm{h})\, \d L_\mathrm{bol} =\\
        &=\,\frac{f_{\rm on,UV}}{\sqrt{2\pi}\sigma}\,\exp\left(\frac{(\log_{10} L_\mathrm{bol} - \log_{10} L_\mathrm{c}(M_\mathrm{h}))^2}{2\sigma^2}\right) \d \log_{10} L_\mathrm{bol} .
        \label{eq:clf_log_normal}
\end{split}
\end{equation}
We assume a power-law $L_\mathrm{c}(M_\mathrm{h})$ relation, parametrized by a slope, $\gamma$, and a normalization $L_{\rm ref}$. In terms of logarithmic quantities:
\begin{equation}
    \log_{10} L_\mathrm{c}(M_\mathrm{h}) = \log_{10} L_{\rm ref} +\gamma\, \left(\log_{10} M_\mathrm{h} - \log_{10} M_{\rm ref}\right), \label{eq:L-M_relation}
\end{equation}
with $M_{\rm ref}$ fixed to $\log_{10} M_{\rm ref}/\msun = 12.5$.

By fitting 
the quasar clustering and the QLF at any given redshift, we have enough information to constrain the quasar luminosity-halo mass relation ($\gamma$ and $L_{\rm ref}$), its intrinsic scatter ($\sigma$), and the active fraction of quasars ($f_{\rm on,UV}$) -- see Table \ref{tab:clf_parameters}. Once these quantities are known, the QHMF can be obtained by statistically assigning quasars to subhalos and selecting only the subhalos whose quasars are brighter than some luminosity threshold, $L_\mathrm{thr}$ (which is usually set according to observations):
\begin{equation}
    \mathrm{QHMF}( M_\mathrm{h}|L_\mathrm{bol}>L_{\rm thr}) = \mathrm{HMF}(M_\mathrm{h})\,\int_{ L_{\rm thr}}^\infty {\rm CLF}( L_\mathrm{bol}| M_\mathrm{h}) \, \d  L_\mathrm{bol} ,\label{eq:qhmf}
\end{equation}
where HMF stands for the (sub)halo mass function.
A comparison between the QHMF and the HMF can then return the value of the UV-luminous quasar duty cycle, $\varepsilon_\mathrm{QSO}$:
\begin{equation}
\begin{split}
        \varepsilon_{\rm QSO} 
        = \frac{\int_{M_{\rm med}}^{\infty} \mathrm{QHMF}( M|L_\mathrm{bol}>L_{\rm thr})\,\d M}{\int_{M_{\rm med}}^{\infty} \mathrm{HMF}( M)\,\d M}.\\
        \label{eq:duty_cycle}
        \end{split}
\end{equation}
The lower integration limit is set to the median value\footnote{The median of the QHMF is defined as the halo mass $M_\mathrm{med}$ satisfying the relation $
   \int_{M_{\rm med}}^{\infty}{\rm QHMF}(M_\mathrm{h}) = 0.5\,\int_{0}^{\infty} {\rm QHMF}(M_\mathrm{h})$.} of the QHMF, 
$M_\mathrm{med}$ (see, e.g., \citealt{ren_trenti2020}).
For more details on the parametrization employed for the CLF and the definition of the various quantities at play, we refer the reader to Sec. 2 in \citet{pizzati2024a} and Sec. 2 and Appendix A in \citet{pizzati2024b}. 

The framework developed in these works builds on large-volume, dark-matter-only cosmological simulations. In particular, \citet{pizzati2024b} uses the new FLAMINGO-10k simulation (part of the FLAMINGO project, \citealt{flamingo,FlamingoII}), which evolves $10080^3$ cold dark matter (CDM) particles and $5600^3$ neutrino particles in a box size of $L=2.8\,\cGpc$ assuming the ``3x2pt + all'' cosmology from \citet{abbott_des2022}\footnote{The cosmology parameters are: $\Omega_\mathrm{m} = 0.306$, $\Omega_\mathrm{b} = 0.0486$, $\sigma_8 = 0.807$, $\mathrm{H}_0 = 68.1\,\kms\,{\rm Mpc}^{-1}$, $n_\mathrm{s} = 0.967$; the summed neutrino mass is $0.06\,\mathrm{eV}$.}. The model includes subhalos, which are found using the upgraded Hierachical Bound-Tracing (HBT+) code \citep[][]{hbt, hbt_plus}{}{}. Subhalo masses, $M_\mathrm{h}$, are defined as peak bound masses\footnote{In practice, we compute the mass of each (sub)halo by summing up the mass of all its bound particles and consider the largest mass that a (sub)halo has had across cosmic history.}.

In the analysis performed in \citet{pizzati2024a}, we applied this framework to the quasar 
auto-correlation functions measured by \citet{eftekharzadeh2015} ($z\approx2.5$) and \citet{shen2007} ($z\approx4$) 
using wide-field spectroscopic surveys such as SDSS \citep[][]{york2000} and BOSS \citep[][]{ross2013}. In particular, we 
showed that the $z\approx4$ clustering measurements of \citet{shen2007} imply a characteristic host halo mass for quasars 
of $\log_{10} M_\mathrm{h}/\msun \approx 13.3$, corresponding to a very large UV-luminous quasar duty cycle of 
$\varepsilon_\mathrm{QSO}=33^{+34}_{-23}\,\%$.
In \citet{pizzati2024b}, we extended the framework to 
interpret the quasar-galaxy cross-correlation function recently measured by \citet{eilers2024} at $z=6.25$. This work 
exploited the JWST NIRCam wide-field slitless spectroscopic mode to pick up \OIII emitting galaxies in quasars fields, and 
inferred the clustering of quasars by measuring the cross-correlation function between quasars and \OIII emitting galaxies 
in conjunction with the auto-correlation function of these galaxies. By simultaneously fitting these two quantities, 
\citet{pizzati2024b} 
found a characteristic host mass for $z\approx6$ quasars of $\log_{10} M_\mathrm{h}/\msun \approx 12.5$, lower than 
the one found at $z\approx4$ and in line with results at $z\approx2.5$. 

\begin{table}
\centering
\setlength{\extrarowheight}{3pt}
\caption{Constraints (median values and 16th-84th percentiles) on the parameters describing the conditional luminosity function (CLF; eq. \ref{eq:clf_log_normal}) of quasars at $z\approx4$ and $z\approx6$. Taken from \citet{pizzati2024a,pizzati2024b}.}

\begin{tabular}{c | c c c c }
\toprule
 Redshift & $\sigma$ & $\log_{10} L_{\rm ref}$ [$\ergs$]    & $\gamma$  & $f_{\rm on}$ [\%]\\ 
\midrule
$z\approx4$ &
$0.20^{+0.13}_{-0.08}$ &
$45.2^{+0.3}_{-0.3}$     &   
$2.00^{+0.22}_{-0.23}$ &
$51^{+32}_{-31}$ \\
\midrule
$z\approx6$ &
$ 0.55_{-0.31}^{+0.37} $ &
$ 46.45_{-1.35}^{+0.79} $      &   
 $ 3.17_{-0.34}^{+0.32} $  &
 $ 3.9_{-3.2}^{+21} $ \\
\bottomrule
\end{tabular}
\label{tab:clf_parameters}
\end{table}

However, when converting these host halo masses into peak heights\footnote{The peak height $\nu(M_\mathrm{h},z)$ is formally 
defined as $\nu(M_\mathrm{h},z) = \delta_\mathrm{c}/\sigma(M_\mathrm{h},z)$ -- with $\delta_\mathrm{c}\approx 1.69$ being the critical linear density for 
spherical collapse and $\sigma^2(M_\mathrm{h},z)$ the variance of the linear density field smoothed on a scale 
$R(M_\mathrm{h})$; we compute $\nu(M_\mathrm{h},z)$ using the python 
package \code{colossus} \citep[][see Sec. 5 in \citealt{pizzati2024b}]{colossus_diemer2018}.}, $\nu(M_\mathrm{h})$ -- which measure 
how rare the large-scale over-density fluctuations are in the original linear field -- we find that quasar clustering measurements 
at $z\approx4$ and $z\approx6$ point to similar values of $\nu\approx4-6$. This implies that high-$z$, UV-luminous quasars 
seem to live in similarly biased and over-dense environments, corresponding to $(4-6)\sigma$ peaks in the initial linear 
density field \citep[see also, e.g.,][]{costa2024}. Due to the rapid decline of the unobscured quasar number density with 
redshift (solid gray line in the right panel of Fig. \ref{fig:overview}), these similar environments lead to very different 
values for the quasar UV-luminous duty cycles at $z\approx4$ and $z\approx6$: while UV-luminous $z\approx4$ quasars 
are sufficiently abundant to occupy a 
large fraction of the coeval $\nu\approx4-6$ halos, at $z\approx6$ quasars are so rare that the 
same occupation fraction drops by more than an order of magnitude, with an implied duty cycle of $\varepsilon_\mathrm{QSO}=0.9^{+2.3}_{-0.7}\%$. 

We report the inferred values of the parameters describing the CLF and the $L_\mathrm{c}(M)$ relation (eq. \ref{eq:clf_log_normal}-\ref{eq:L-M_relation}) at $z\approx4$ and $z\approx6$ in Table \ref{tab:clf_parameters}. Further discussion on the comparison between quasar clustering results at these two redshifts can be found in 
Sec. 5 of \citet{pizzati2024b} (see also \citealt{eilers2024}). 
We mention the caveat, however, that the strong clustering measured at $z\approx4$ is rather surprising and it is yet to be fully accounted for by any evolutionary models of quasar activity \citep[][and references therein]{pizzati2024a}. Additionally, several other studies \citep[e.g.,][]{timlin2018,he2018,garcia-vergara2019}{}{} have also attempted to measure quasar clustering at $z\approx4$, challenging the exceptionally strong clustering inferred by \citet{shen2007}. Nevertheless, the \citet{shen2007} measurement remains the most robust, as it is based on a large sample of spectroscopically-selected quasars. Future spectroscopic surveys (such as DESI, \citealt{yang2023_desi}) will further refine these measurements and provide more stringent constraints on the quasar auto-correlation function up to $z\approx5$. Here, we take the \citet{shen2007} result at face value, but stress the fact that our conclusions for $z\approx4$ and $z\approx6$ are completely independent. 

\begin{table*}
\centering
\setlength{\extrarowheight}{3pt}
\caption{Constraints (median values and 16th-84th percentiles) on the UV-luminous active fraction $f_\mathrm{on,UV}$ (coming from clustering measurements; see \citealt{pizzati2024a,pizzati2024b}) and on the obscured:unobscured ratio for LRDs, $r_\mathrm{obsc}$ (from abundance arguments; see Sec. \ref{sec:lrd_abundance}), at $z=4$ and $z=6.25$. The product $f_\mathrm{on,UV}r_\mathrm{obsc}$ exceeds unity at both redshifts, which is unphysical. In the last columns, we also report the median mass, $M_\mathrm{med}^\mathrm{(faint)}$ for the halos hosting faint unobscured quasars (see Fig. \ref{fig:qhmfs}) and the number density of halos above this mass, $n_\mathrm{h}(>M_\mathrm{med}^\mathrm{(faint)})$, to be compared with the LRD number density of $\Phi_\mathrm{LRD}\approx1.3\times10^{-5}\,\cMpc^{-3}$.}
\begin{tabular}{c | c c c c | c c c }
\toprule
 Redshift  & $\log_{10} f_{\rm on, UV}$ & $\log_{10} r_\mathrm{obsc}$
 & $\log_{10} f_\mathrm{on,UV}\cdot
 r_\mathrm{obsc}$
 & $f_\mathrm{on,UV}\cdot
 r_\mathrm{obsc}$
 & $\log_{10} M_\mathrm{med}^\mathrm{(faint)}/\msun$ 
 & $n_\mathrm{h}(>M_\mathrm{med}^\mathrm{(faint)})$
 & $\Phi_\mathrm{LRD}/n_\mathrm{h}(>M_\mathrm{med}^\mathrm{(faint)})$\\ 
\midrule
$z=6.25$ &
$ -1.40_{-0.74}^{+0.83} $ &
$ 2.9\pm0.5 $      &   
$ 1.5_{-0.9}^{+1.0} $  &
$ 32_{-28}^{+284} $ &
$\approx 12.15$ &
$\approx 7.2\times10^{-7}\,\cMpc^{-3}$ &
$\approx 18$ \\
\midrule
$z=4$ &
$ -0.29_{-0.41}^{+0.21} $  &
$ 1.3\pm0.3 $      &   
$ 1.0_{-0.5}^{+0.4} $ &
$ 10_{-7}^{+15} $ &
$\approx 12.76$ &
$\approx 1.6\times10^{-6}\,\cMpc^{-3}$ &
$\approx 8.1$ \\
\bottomrule
\end{tabular}
\label{tab:lrd_vs_qso}
\end{table*}

In Fig. \ref{fig:qhmfs}, we show the QHMFs obtained by our model at $z\approx4$ and $z\approx6$, together with 
HMFs at the respective redshifts. As discussed above, the QHMF can be obtained only 
once a bolometric luminosity threshold for quasars has been set. Both quasar clustering measurements on which our work is 
based \citep[][]{shen2007, eilers2024} focus on very bright unobscured quasars with $L_\mathrm{bol}\approx10^{47}\,\ergs$, with 
the work of \citet{shen2007} extending down to slightly fainter objects of $L_\mathrm{bol}>10^{46.7}\,\ergs$. For 
consistency (see also Appendix D of \citealt{pizzati2024b}), we show
our $z\approx6$ QHMF results setting the same bolometric 
luminosity threshold employed by \citet{shen2007} at $z\approx4$ (i.e., $L_\mathrm{bol}=10^{46.7}\,\ergs$). The QHMFs obtained in this way 
are plotted in Fig. \ref{fig:qhmfs} with red ($z\approx6$) and green ($z\approx4$) lines, 
and labeled as ``bright quasars'' as they only refer to the bright end of the unobscured quasar population. 

Fainter, unobscured quasars are found at both $z\approx4$ and $z\approx6$ down to 
$L_\mathrm{bol}\approx10^{45.3}\,\ergs$ \citep[][]{akiyama2018,kulkarni2019,matsuoka2022}. However, 
the clustering of 
this fainter population is still largely unconstrained in the high-$z$ Universe. A first attempt at measuring the 
clustering of $z\approx6$ faint quasars was made by \citet{arita2023}: despite the large uncertainties at play, these authors 
find a relatively large characteristic host halo mass of $M_\mathrm{h}=7^{+11}_{-6}\times10^{12}\,\msun$ 
(but see Appendix C of \citealt{pizzati2024b}, where it is shown that different assumptions on the quasar correlation function make these constraints much weaker). The relatively large inferred host mass 
for the 
faint quasar 
population would be in line with results at lower redshift ($z\lesssim2.5$), which 
generally predict little to no dependence of quasar clustering with bolometric 
luminosity \citep[e.g.,][]{shen2009, eftekharzadeh2015}.

As our model is based on an empirical relation between quasar luminosities and (sub)halo masses, it can be used to 
predict the clustering of faint unobscured quasars at high redshift. 
With light-colored lines in Fig. \ref{fig:qhmfs}, we plot the predictions for the ``faint quasars'' QHMFs at the two redshifts of interest. 
These QHMFs are obtained by lowering the bolometric luminosity threshold, $L_\mathrm{thr}$ in eq. \ref{eq:qhmf}, down 
to $L_\mathrm{bol}=10^{45.5}\,\ergs$. We note that such a low bolometric luminosity threshold implies that the results are sensitive to the relation between \textit{faint} quasar luminosities and host halo masses. This 
relation is based on the extrapolation of our CLF parametrization down to low $L_\mathrm{bol}$, and it currently lacks 
support by constraints on the clustering of faint unobscured quasars. However, our fitting framework matches the unobscured QLF over the entire range of magnitudes, from the 
very bright to the very faint end, with a minimal number of parameters.  Therefore, while faint quasar clustering studies will ultimately
 test our predictions,
 the QHMFs shown in Fig. \ref{fig:qhmfs} for faint quasars represent our best knowledge of how faint quasars 
 populate the host halo mass spectrum, and are informed by our current understanding of unobscured quasar demographics.

At $z\approx6$ (left panel of Fig. \ref{fig:qhmfs}), we predict that the ``faint quasars'' QHMF peaks 
at $\log_{10} M_\mathrm{h}/\msun\approx12.15$, with a rather large spread in the host mass distribution 
(0.5 dex at 1 standard deviation). 
This implies a very mild dependence of clustering on bolometric luminosity, as a change of $\approx1$ dex in $L_\mathrm{bol}$ results in a change of $\approx0.3$ dex in the median of the host mass distribution, $M_\mathrm{med}$.
This mild dependence 
is driven by two factors: a steep $L_\mathrm{bol}-M_\mathrm{h}$ relation and a large scatter around this relation\footnote{The slope of the $L_\mathrm{bol}-M_\mathrm{h}$ relation and its scatter are directly constrained by a combination of the quasar clustering strength and the \textit{shape} of the quasar luminosity function \citep[][]{pizzati2024a}.}. These results are in 
broad agreement with clustering studies at low redshift, which find little to no dependence of clustering strength on luminosity 
\citep[][]{croom2005,myers2006,shen2009} and attribute that to a large scatter in quasar luminosities at 
fixed halo mass \citep[e.g.,][]{adelberger_steidel2005,lidz2006}. 

The strong clustering measured for bright quasars at $z\approx4$ implies a slightly different dependence 
of quasar clustering on luminosity, with $\approx1$ dex in $L_\mathrm{bol}$ corresponding to $\approx0.5$ dex in $M_\mathrm{med}$. 
Such a 
luminosity dependence is a consequence of the large duty cycle measured for bright quasars: if these quasars occupy 
a large fraction of the available massive halos, fainter quasars will inevitably need to reside in less massive hosts. 
In practice, this is achieved in our model with a small predicted scatter for the $z\approx4$ $L_\mathrm{bol}-M_\mathrm{h}$ 
relation \citep[also found by][]{white_2008, wyithe_loeb2009,shankar_2010}. 
The slightly different dependence of clustering on luminosity at the two redshifts considered, while interesting, has little impact on the conclusions presented in this work: at both redshifts, faint quasars also live in massive 
halos corresponding to highly biased environments which
trace back to rare $\approx4\sigma$ 
fluctuations in the linear density field.

\subsection{Connecting the UV-luminous duty cycle to the AGN obscured population}

Having described current constraints on the duty cycle and host mass distribution of UV-luminous, unobscured quasars, we turn our attention to the large population of LRDs/obscured AGN discussed in Sec. \ref{sec:lrd_abundance}. The most general question connected to this obscured high-$z$ population is how it fits into our understanding of SMBH accretion/AGN activity across the history of the Universe. In this context, determining whether LRDs and UV-selected quasars are drawn from the same population of dark matter halos can offer key insights into the nature of these sources. According to AGN unification models \citep[e.g.,][]{Antonucci93, padovani2017}, the diversity of AGN emission across the electromagnetic spectrum can be entirely explained by a viewing-angle effect: the intrinsic emission from a quasar/AGN varies for different lines of sight because of, e.g., dust and gas obscuration. The natural consequence of this model is that all types of AGN (irrespective of their observed SEDs) share the same \textit{intrinsic} properties, such as the bolometric luminosity, SMBH mass, and host halo mass distributions. Hence, if LRDs fit into this AGN unification picture, we expect them to reside in the same halos as comparably luminous UV-selected quasars. However, several studies at low $z$ have challenged this AGN unification scenario by showing that obscured (type-2 or reddened type-1) quasars live in different dark matter halos than those of UV-luminous, type-1 quasars \citep[e.g.,][]{hickox2011,allevato2014, petter2023, cordova_rosado2024}. According to these studies, obscured quasars/AGN represent a different stage in the co-evolution between accreting SMBHs and their host galaxies/halos. Analogously, LRDs could also represent a different evolutionary phase in the accretion history of SMBHs. If that is the case, the host halo mass distribution of LRDs could be different than the one of unobscured quasars, even when matching their bolometric luminosities and SMBH masses. An obvious consequence of this hypothesis is that LRDs would be described by very different scaling relations (e.g., SMBH mass-halo/galaxy mass) than those in place for UV-luminous quasars, as identical SMBH masses would correspond to very different host halo/galaxy masses. 

In this work, we point out that an indirect answer to whether LRDs and UV-selected quasars reside in the same dark matter halos comes from current constraints on the clustering of quasars at $z\approx4-6$ (Sec. \ref{sec:unobscured_quasars}). From these constraints, we conclude that LRDs and unobscured quasars cannot be drawn from the same host halo distribution. Hence, their different SED properties reveal fundamental differences in their scaling relations. 
Our argument is simple: clustering measurements determine the host mass distribution of unobscured quasars; if LRDs followed the same distribution, the large obscured fraction derived in Sec. \ref{sec:lrd_abundance} implies that LCDM cosmology would not produce enough halos at these masses to accommodate this abundant population. 

The argument can be visualized in Fig. \ref{fig:qhmfs}: using dark (light) purple lines, we show the QHMFs of bright (faint) quasars scaled up by the obscured:unobscured ratios, $r_\mathrm{obsc}$, determined in Sec. \ref{sec:lrd_abundance} (plotted with colored arrows for reference). These obscured ratios are independent of bolometric luminosities, and increase rapidly with redshift from $r_\mathrm{obsc}\approx20^{+20}_{-10}:1$ at $z=4$ to $r_\mathrm{obsc}\approx815^{+1600}_{-545}:1$ at $z=6.25$. 
By multiplying the QHMF by $r_\mathrm{obsc}$,
we are effectively computing the host mass distribution for LRDs/obscured AGN under the hypothesis that they reside in the same kind of halos as UV-luminous quasars. 
At both $z=6.25$ (left panel in Fig. \ref{fig:qhmfs}) and $z=4$ (right), the host halo mass distributions for LRDs exceed the respective halo mass functions (HMFs). This is unphysical: cosmology sets hard (and well-constrained) limits on the number of (sub)halos that are available as quasar hosts as a function of mass, and the LRD number densities appear to be incompatible with these limits\footnote{Note that this argument is valid only for a maximum occupation fraction of unity. Given that we model the distribution of all subhalos, however, it is natural to assume that each subhalo can host at most one accreting SMBH at its center.}.

We can quantify this by considering the UV-luminous active fraction, $f_\mathrm{on,UV}$, which is a parameter in our CLF model (see eq. \ref{eq:clf_log_normal}) and is closely related to the UV-luminous duty cycle \citep[][]{pizzati2024a}. The parameter $f_\mathrm{on,UV}$ represents the fraction of SMBHs that are actively accreting and unobscured at the same time. If we multiply this UV-luminous active fraction by the obscured:unobscured ratio $r_\mathrm{obsc}$, 
we are effectively computing an ``obscured'' active fraction (i.e., the fraction of halos hosting actively accreting LRDs/obscured AGN).
The physical limit set by the number of available sub(halo) hosts can be then rephrased as $f_\mathrm{on,UV}\cdot r_\mathrm{obsc}<1$. 
In Table \ref{tab:lrd_vs_qso}, we report the values of $f_\mathrm{on,UV}$ and $r_\mathrm{obsc}$ and of their product at the two redshifts of interest, $z=6.25$ and $z=4$. We find that, despite the large uncertainties at play, these products are significantly larger than unity, with a value of $f_\mathrm{on,UV}\cdot r_\mathrm{obsc}\approx10$ at $z=4$ and $f_\mathrm{on,UV}\cdot r_\mathrm{obsc}\approx36$ at $z=6.25$. Coming back to the visual representation in Fig. \ref{fig:qhmfs}, the product $f_\mathrm{on,UV} \cdot r_\mathrm{obsc}$ represents the maximum ratio between the scaled-up QHMFs (purple lines; see also eq. \ref{eq:qhmf}) and the HMFs (dashed lines).

An even simpler way to present this argument is to consider the median mass values, $M_\mathrm{med}$, for, e.g., the faint-quasar QMHFs (Fig. \ref{fig:qhmfs}, light-colored lines). In the right-hand columns of Table \ref{tab:lrd_vs_qso}, we report these  $M_\mathrm{med}^\mathrm{(faint)}$ values at the two redshifts of interest, together with the number density of halos above these mass thresholds, $n_\mathrm{h}(>M_\mathrm{med}^\mathrm{(faint)})$. When compared to the number density of LRDs, $\Phi_\mathrm{LRD}\approx1.3\times10^{-5}\,\cMpc^{-3}$ (which is approximately constant with redshift; see Sec. \ref{sec:lrd_abundance}), these number densities are a factor of $\approx8.1$ ($\approx18$) smaller at $z=4$ ($z=6.25$). This confirms the fact that LRDs are simply too numerous to live in the same (sub)halos as UV-luminous quasars. As discussed in Sec. \ref{sec:unobscured_quasars}, these halo masses correspond to similar environments at $z\approx4$ and $z\approx6$ (i.e., $(4-6)\sigma$ peaks in the linear density field; Fig. \ref{fig:qhmfs}). Since the number density of these environments 
is roughly constant with redshift \citep[e.g.,][]{tinker2008} (and so is $\Phi_\mathrm{LRD}$), LRDs outnumber their candidate host halos by similar factors at the two redshifts considered.

As a final note, we point out that our results are valid for any values of the quasar bolometric luminosities. Yet, in Sec. \ref{sec:lrd_abundance} and \ref{sec:unobscured_quasars} we considered bright ($L_\mathrm{bol}>10^{46.7}\,\ergs$) and faint ($10^{45.5}\,\ergs<L_\mathrm{bol}<10^{46.7}\,\ergs$) quasars separately because their properties are constrained differently. In particular, the abundance of obscured AGN is better constrained at faint bolometric luminosities by the large sample of LRDs with $L_\mathrm{bol}\approx10^{46}\,\ergs$; the evidence for an analog obscured population at large bolometric luminosities is instead only tentative (Sec. \ref{sec:lrd_abundance}). On the other hand, the clustering of bright unobscured quasars has been directly measured (Sec. \ref{sec:clustering_duty_cycle}), but the QHMF and duty cycle for the faint quasar population is solely based on the extrapolation of our model to fainter bolometric luminosities -- which constrained to match the faint end of the QLF. For this reason, the results presented lead to different conclusions depending on the bolometric luminosities considered. If a large obscured population is indeed present at $L_\mathrm{bol}\approx10^{47}\,\ergs$, then this is already in direct conflict with 
with constraints on the host masses and duty cycle of bright unobscured quasars \citep[][]{shen2007, eilers2024}. A measurement of quasar clustering at the faint end of the QLF ($L_\mathrm{bol}\gtrsim10^{45.5}\,\ergs$), on the other hand, would provide support for our predictions for the properties of faint unobscured quasars, and will make it possible to directly compare the properties of UV-luminous quasars and LRDs at the same bolometric luminosities.

\section{The host mass and duty cycle of little red dots: a mock analysis} \label{sec:lrd_mocks}

\begin{figure*}
	\centering
	\includegraphics[width=0.95\textwidth]{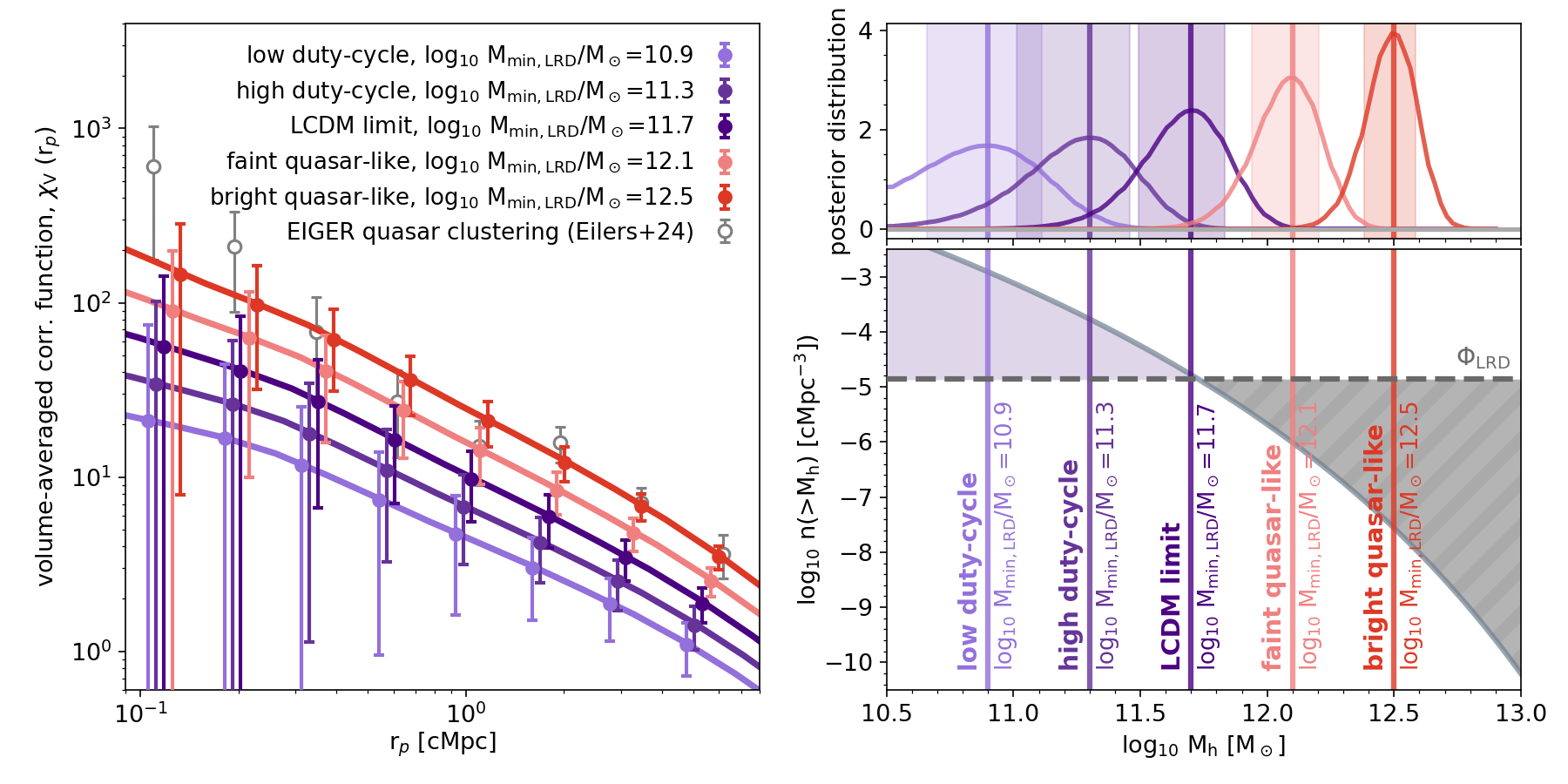}
	 \caption{\textit{Left:} Mock measurements (colored data points) for the LRD-galaxy cross-correlation functions obtained for different values of the minimum host mass for LRDs, $M_\mathrm{min,LRD}$. The measurements are obtained by putting together 10 different LRD fields, and extracting galaxy counts by setting a minimum host mass for galaxies (i.e., \OIII emitters) of $M_\mathrm{min,OIII}=10^{10.56}\,\msun$ and a background galaxy number density of $n_\mathrm{OIII}=7.84\times10^{-4}\,\cMpc^{-3}$. The theoretical predictions for these cross-correlation functions are coming from the model of \citet{pizzati2024b} and are shown as solid colored lines. Error bars are computed by assuming Poisson uncertainties on the galaxy number counts. Gray points refer to the UV-luminous quasar-galaxy cross-correlation function measurements from the EIGER survey \citep{eilers2024}.  \textit{Top right:} Mock inference analysis for the LRD-galaxy cross-correlation function measurements, as a function of the minimum host LRD mass, $M_\mathrm{min,LRD}$. Values of $M_\mathrm{min,LRD}$ considered for the mock measurements are color-coded as in the other panels. The posterior distributions are obtained by computing the agreement between the mock measurements and the theoretical models for different minimum host LRD masses. Shaded regions show the 16th and 84th percentiles of their respective posterior distributions. 
     \textit{Bottom right:} Number density of $z=6.25$ halos above $M_\mathrm{h}$, $n_\mathrm{h}(>M_\mathrm{h})$, as a function of halo mass $M_\mathrm{h}$ (solid gray line). The values of $M_\mathrm{min,LRD}$ considered in the analysis are highlighted with colored vertical lines. The dashed horizontal line corresponds to the LRD number density, $\Phi_\mathrm{LRD}$. The purple (gray) shaded area shows the region for which $M_\mathrm{h}<10^{11.7}\,\msun$ ($M_\mathrm{h}>10^{11.7}\,\msun$). In the purple region, $\Phi_\mathrm{LRD}<n_\mathrm{h}(>M_\mathrm{h})$ and hence the number of LRDs is less than the number of host halos available, whereas the gray region is unphysical as LRDs are too abundant for the number of host (sub)halos (assuming a maximum occupation fraction of unity).  
    \label{fig:mocks}
 	}
\end{figure*}

The indirect arguments presented in the previous section 
suggest that LRDs cannot live in the same dark matter halos 
as unobscured UV-luminous quasars, and hence -- provided their bolometric luminosities are correctly estimated -- they may 
constitute a fundamentally different population of accreting SMBHs.
How do we determine this new population's host halo masses and duty cycle? 
In this section, 
we argue that this can be done using current (and upcoming) JWST observations. 

Existing JWST programs such as EIGER \citep[][]{kashino2022, eilers2024} and ASPIRE \citep[][]{aspire_wang2023} have already 
shown that the clustering of luminous, UV-selected quasars can be effectively measured 
using JWST NIRCam slitless spectroscopy to study the 
distribution of \OIII line emitting galaxies in the neighboring regions of the quasars. The same strategy can be applied to any other population of objects: the cross-correlation between this population and \OIII line 
emitters at a certain redshift can be measured, and the clustering of this said population can be inferred by simultaneously 
constraining the auto-correlation function of the \OIII line emitters.

In the following, we examine a simple proof-of-concept analysis that aims to measure the clustering of LRDs using JWST\footnote{An alternative approach would be to directly measure the auto-correlation function of LRDs. Even though LRDs have a relatively high number density, however, measuring an autocorrelation function would require very large samples that are challenging to obtain given the small FoV of JWST.}. We focus here on $z=6.25$, which is the redshift at which the clustering of UV-luminous quasars with \OIII emitters has already been measured by the EIGER survey \citep[][see also Sec. \ref{sec:unobscured_quasars}]{eilers2024}.  
Following \citet{eilers2024} (see also, e.g., \citealt{kaiser1984,martini2001,haiman_hui2001}), 
we postulate that LRDs inhabit a fraction of all the (sub)halos whose mass is larger than some minimum mass threshold, $M_\mathrm{min,LRD}$\footnote{In other words, we do not model the LRD host mass distribution parametrically as described in Sec. \ref{sec:clustering_duty_cycle} for unobscured quasars, but we assume that such a distribution can be obtained by rescaling the HMF above the minimum mass threshold $M_\mathrm{min,LRD}$. A more sophisticated parametrization would result in large degeneracies in the parameter space that could not be resolved by clustering measurements alone \citep[e.g.,][]{pizzati2024a,munoz2023}.}. This fraction is equal to the LRD duty cycle, $\varepsilon_\mathrm{LRD}$, and can be determined by comparing the LRD number density ($\Phi_\mathrm{LRD}$ in Sec. \ref{sec:lrd_abundance}) to the abundance of halos with $M_\mathrm{h}>M_\mathrm{min}$. We note that we only consider LRDs with quasar-like bolometric luminosities (i.e., with the same bolometric luminosities as faint $z\approx6$ quasars, $L_\mathrm{bol}>10^{45.5}\,\ergs$), 
as we are interested in matching LRDs and UV-luminous quasars in $L_\mathrm{bol}$ space.

We consider five different values of the minimum host (sub)halo mass for LRDs: $\log_{10} M_\mathrm{min,LRD}/\msun = 10.9, 11.3, 11.7, 12.1, 12.5$. In the bottom right panel of Fig. \ref{fig:mocks}, we put these values into context by showing the number density of $z=6.25$ halos above $M_\mathrm{h}$, $n_\mathrm{h}(>M_\mathrm{h})$, as a function of halo mass (solid grey line); we highlight the values of $M_\mathrm{min,LRD}$ considered with
colored vertical lines. By comparing the LRD number density ($\Phi_\mathrm{LRD}$, dashed horizontal line) to the integrated halo mass function ($n_\mathrm{h}(>M_\mathrm{h})$) for different minimum halo masses, we can directly relate the abundance of LRDs to that of available host dark matter halos. 
We find that the number of LRDs equals the number of host halos (i.e., the duty cycle is equal to unity) for a minimum host mass of $M_\mathrm{min,LRD}\approx10^{11.7}\,\msun$. Assuming that there can be only one LRD per (sub)halo, values of $M_\mathrm{min,LRD}$ above this threshold mass are unphysical. Values significantly lower than this threshold, on the other hand, imply a low duty cycle for LRDs, as only very few (sub)halos host LRDs at any given time. 

Based on this discussion, we refer to the five different $M_\mathrm{min,LRD}$ cases considered in the following way (see Fig. \ref{fig:mocks}): ``low duty-cycle'' ($M_\mathrm{min,LRD}/\msun = 10^{10.9}\,\msun$), ``high duty-cycle'' ($M_\mathrm{min,LRD}/\msun = 10^{11.3}\,\msun$), ``LCDM limit'' ($M_\mathrm{min,LRD}/\msun = 10^{11.3}\,\msun$), ``faint quasar-like'' ($M_\mathrm{min,LRD}/\msun = 10^{12}\,\msun$), ``bright quasar-like'' ($M_\mathrm{min,LRD}/\msun = 10^{12.5}\,\msun$). 
The first case (``low duty-cycle'') corresponds to a duty cycle of $\varepsilon_\mathrm{LRD}\approx1\%$, which is close to the duty cycle measured by \citet{pizzati2024b} for UV-luminous quasar activity at the same redshift. In the second case, the implied LRD duty cycle increases to $\varepsilon_\mathrm{LRD}\approx10\%$. The third case corresponds to the physical limit of a duty cycle of $\approx100\%$. The last two cases, instead, would imply a duty cycle above unity, and correspond to host masses characteristic of UV-luminous quasars. Based on the discussion of Sec. \ref{sec:unobscured_quasars}, we associate the case $M_\mathrm{min,LRD}=10^{12.1}\,\msun$ to faint ($L_\mathrm{bol}\gtrsim10^{45.5}\,\ergs$) quasars -- which have the same $L_\mathrm{bol}$ as LRDs --, 
while the larger mass of $M_\mathrm{min,LRD}=10^{12.5}\,\msun$ is close to the one found for luminous ($L_\mathrm{bol}\approx10^{47}\,\ergs$) unobscured quasars by \citet{eilers2024}. 

The question we want to address here is whether we can use clustering measurements based on JWST slitless spectroscopy data to distinguish between these different $M_\mathrm{min,LRD}$ cases. We consider the following mock setup: JWST/NIRCam grism is used to image 10 different LRD fields. The distribution of \OIII line emitters in these fields can be employed to measure a LRD-galaxy cross-correlation function, from which the host mass and duty cycle of LRDs can be determined by exploiting the constraints on the galaxy-galaxy auto-correlation function (\citealt{eilers2024}, Huang et al., in prep.).

In practice, we use the framework developed in \citet{pizzati2024b}, which outputs the cross-correlation function of any populations of objects that are tracers of the underlying distribution of dark matter halos\footnote{We use the FLAMINGO-10k large-volume cosmological simulation (Sec. \ref{sec:clustering_duty_cycle}) to build an analytical model for the cross-correlation function of any sets of halos with masses $M_j$ and $M_k$, $\xi_h(M_j, M_k; r)$. An appropriate weighting scheme can then return the cross-correlation function between two different halo tracer populations. For more details on the model and the cosmological simulation employed, we refer the reader to \citet{pizzati2024b}.}. We employ this model to predict the LRD-galaxy cross-correlations for the different values of $M_\mathrm{min,LRD}$. \OIII line emitters are assumed to live in halos with a fixed threshold mass of $M_\mathrm{min,OIII}=10^{10.56}\,\msun$, which is set according to the results of \citet{eilers2024} (see also Huang et al., in prep.). Based on these cross-correlation functions, we generate mock measurements by computing the expected number of galaxies as a function of the projected distance in each LRD field. 
The expected galaxy counts are obtained by setting a background galaxy number density of $n_\mathrm{OIII}=7.84\times10^{-4}\,\cMpc^{-3}$, which is obtained by integrating the \OIII emitter luminosity function of \citet{matthee2023} down to the threshold luminosity of $L_\mathrm{OIII,5008}=10^{42}\,\ergs$. 
We put together the 10 mock LRD fields 
and we compute the volume-averaged cross-correlation function, $\chi_V$, by projecting the galaxy 3-d distributions over a comoving distance of $\pi_\mathrm{max}=9.8\,\cMpc$, corresponding to a line-of-sight velocity of $1000\,\kms$ at the redshift considered.

In the left panel of Fig. \ref{fig:mocks}, we show the mock LRD-galaxy cross-correlation functions for different values of $M_\mathrm{min,LRD}$. We also show for reference the UV-luminous quasar-galaxy cross-correlation function measured by \citet{eilers2024} by putting together 4 different quasars fields from the EIGER survey \citep[][]{kashino2022}. We note that, as also done in \citet{eilers2024},
the error bars we show are computed by considering only the contribution of Poisson uncertainties on the number counts. Other contributions to the error budget, such as cosmic variance or possible correlations between different data points, are neglected in this work and will be analyzed in a forthcoming study (Huang et al., in prep.). 

The precision of our inference analysis is shown in the bottom left panel of Fig. \ref{fig:mocks}.
These posterior distributions are obtained by fitting the mock data with the LRD-galaxy cross-correlation function models obtained by varying the LRD mass threshold parameter, $\log_{10} M_\mathrm{min,LRD}/\msun$. For each of these models, we compute the value of the $\chi^2$ and plot in Fig. \ref{fig:mocks} the quantity $\exp(-\chi^2/2)$ (normalized to unity).  
By looking at the different posterior distributions, we learn that by putting together 10 LRD fields we can already constrain the values of $M_\mathrm{min,LRD}$ (and hence the characteristic host mass of LRDs) with an uncertainty of $\approx0.1-0.3$ in $\log_{10} M_\mathrm{h}$. The posteriors are narrower and more peaked for larger $M_\mathrm{min,LRD}$. This follows from the fact that high-mass halos are more strongly clustered, and hence the clustering signal is stronger for large $M_\mathrm{min,LRD}$ (left panel of Fig. \ref{fig:mocks}). In all cases considered, the uncertainty in $M_\mathrm{min,LRD}$ is small enough that, in principle, it could be possible to tell apart the different scenarios. A larger number of LRD fields would be necessary, however, to reduce the uncertainties on $M_\mathrm{min,LRD}$ even further, and pinpoint its value even for the case of low $M_\mathrm{min,LRD}$.

The discussion presented here shows that, by measuring how galaxies cluster in LRD fields, it is indeed possible to determine whether LRDs live in the same dark matter halos as unobscured quasars (in agreement with, e.g., the AGN unification framework) or whether they are hosted by more common and less-biased environments, as it appears to be necessary given their large number density (see Sec. \ref{sec:clustering_duty_cycle}).  
In this latter case, measuring the host mass distribution of LRDs would also provide a way to quantify their duty cycle ($\varepsilon_\mathrm{LRD}$), which is a fundamental quantity that can help us to shed light on the accretion history of these enigmatic objects. 
A large value of $\varepsilon_\mathrm{LRD}\approx10\%$ would suggest that LRDs have been actively accreting for a large fraction of cosmic time ($\gtrsim100\,\mathrm{Myr}$), and hence -- assuming a standard value for the radiative efficiency -- that they would be able to build the relatively large black hole masses that have been inferred from their broad optical lines (up to $\gtrsim10^8\,\msun$; e.g., \citealt{ greene2024, kocevski2024}). In particular, an accretion timescale of $\gtrsim100\,\mathrm{Myr}$ corresponds to $\gtrsim 2 t_\mathrm{S}$, where $t_\mathrm{S}\approx45\,\mathrm{Myr}$ is the Salpeter time for exponential black hole mass growth \citep[][]{salpeter1964}. This implies that LRDs are detectable above the observational luminosity threshold for at least a few Salpeter times, which is expected if the survey spans about one order of magnitude in luminosity. We point out that, for the same reason, large duty cycles of $\gtrsim50\%$ are not to be expected, because they would imply that almost all LRDs shine above the observational threshold for a time that is much longer than the Salpeter timescale. Considering again a survey spanning about one order of magnitude in luminosity, a standard Eddington-limited growth that remains above the observational threshold for a time $t\gg 2t_\mathrm{S}$ would result in black holes that grow much more than one order of magnitude, and hence end up being more massive than what is actually observed. For this reason, while the threshold mass of $M_\mathrm{min,LRD}\approx10^{11.7}\,\msun$ represents a limit set by cosmological constraints on the number of available (sub)halos, black hole formation physics suggests an even more stringent limit on $M_\mathrm{min,LRD}$: if we require $\varepsilon_\mathrm{LRD}\lesssim30\%$, this implies that $M_\mathrm{min,LRD}$ needs to be lower than $\approx10^{11.5}\,\msun$.

A very low LRD duty cycle $\varepsilon_\mathrm{LRD}\lesssim1\%$, on the other hand, would also be puzzling, as it would raise the question of how to reconcile the large black hole masses measured for LRDs with their inherently sporadic activity. This is the same problem that has been brought up for the high-$z$ UV-luminous quasar population, for which different methods generally infer low values of the quasar duty cycle and/or quasar lifetime \citep[e.g.,][]{khrykin2016, khrykin2019,khrykin2019b, Eilers2018b, Eilers2020, davies2018,Davies2019b,davies2020,worseck2016,worseck2021,Durovcikova2024, eilers2024} that appear to be in tension with their large, $\gtrsim10^9\,\msun$ black hole masses. A possible solution to explain a low value of the duty cycle is super-Eddington accretion: if accretion on black holes takes place in short, radiatively inefficient bursts, then we expect a lower $\varepsilon_\mathrm{LRD}$ because the Salpeter timescale for black hole accretion becomes shorter than $\approx45\,\mathrm{Myr}$. Interestingly, several studies have invoked super-Eddington accretion to explain the puzzling SED features of LRDs \citep[e.g.,][]{greene2024,pacucci_narayan2024, lambrides2024}. Measuring the clustering of LRDs and inferring their duty cycle would provide an independent way to support these claims. 

Finally, if bright LRDs have large black hole masses ($\gtrsim10^8\,\msun$) but live in much smaller halos than UV-selected quasars, they need to obey fundamentally different scaling relations. Constraining the clustering of LRDs would provide insights into these relations: the lower the mass of the host halos, for instance, the more overmassive LRDs need to be with respect to the black hole mass-halo mass relation holding for unobscured quasars. We can also cast this in terms of the black hole mass-stellar mass relation -- which has been extensively discussed in the recent literature \citep[e.g.,][]{pacucci2023_mbhmstar,yue2024_eiger} -- by converting halo masses to stellar masses using the relation provided by \citet{universe_machine}. We find that halo masses in the range $M_\mathrm{h}\approx10^{11}-10^{11.5}\,\msun$ correspond -- at the redshift of interest -- to stellar masses of $M_\star\approx10^{8.4}-10^{9.4}\,\msun$. This implies that, assuming black hole mass measurements are not significantly overestimated, LRDs are highly overmassive with respect to the coeval black hole mass-stellar mass relation, as the ratio between black hole and galaxy masses would be in the range $\approx10\%-100\%$ \citep[see also, e.g.,][]{durodola2024}.

\section{Discussion and summary} \label{sec:conclusions}

In this work, we have examined how the new population of Little Red Dots (LRDs) revealed by JWST compares to the one of UV-luminous quasars that have been studied for decades using wide-field spectroscopic surveys \citep[e.g.][]{fan2022}. 
The basic observational evidence on which our work is based, is that a large fraction of LRDs  
exhibits 
broad emission lines in their spectra, whose properties directly point to the presence of AGN that are (at least partially) responsible for the observed emission \citep[][]{greene2024, kocevski2024}. This, together with their very red colors at optical wavelengths, has led to the interpretation that LRDs could be standard, UV-luminous type-1 quasars whose radiation is (partially) obscured by intervening dust and gas. By correcting for the effects of this obscuration, it is possible to use broad lines to estimate the bolometric luminosities of the SMBHs accreting at the center of LRDs. Several works \citep[e.g.,][]{greene2024, kokorev2024, akins2024} have shown that such (unattenuated) bolometric luminosities extend up to $\approx10^{46}-10^{47}\,\ergs$, well in the range that is characteristic of unobscured, type-1 quasars (Fig. \ref{fig:overview}, left panel). 

Yet, the abundances of LRDs and UV-luminous quasars are remarkably different.
In Fig. \ref{fig:overview}, we have directly compared the redshift evolution for the number density of UV-luminous quasars to the one for LRDs at the same bolometric luminosities. It is well-established that the abundance of unobscured quasars drops exponentially with increasing redshift \citep[e.g.,][]{richards2006, schindler2023}. Spectroscopic \citep[][]{greene2024} and photometric \citep[][]{kokorev2024} surveys of LRDs, instead, find little to no evolution in their number density over a wide redshift range ($z\approx4-8$), with an approximately constant value of $\Phi_\mathrm{LRD}\approx1.3\times10^{-5}\,\cMpc^{-3}$ ($L_\mathrm{bol}>10^{45.5}\,\ergs$). By comparing the number density of LRDs to that of UV-luminous quasars as a function of redshift, we can estimate the obscured fraction of AGN implied by this LRD population. We infer a large and rapidly evolving obscured fraction, ranging from $\approx20:1$ at $z\approx4$ to $\approx2300:1$ at $z\approx7$.

While this obscured fraction is mostly constrained at the bolometric luminosities for which a significant overlap between LRDs and unobscured quasars is present (i.e., $L_\mathrm{bol}\approx10^{45}-10^{46}\,\ergs$), we find tentative evidence for it to extend to even larger bolometric luminosities ($L_\mathrm{bol}\gtrsim10^{47}\,\ergs$). There are two arguments in support of this evidence: (a) photometric observations \citep[][]{kokorev2024} constrain the \textit{shape} of the LRD bolometric luminosity functions to closely resemble that of UV-luminous quasars \citep[][]{niida2020, schindler2023,matsuoka2023}, implying an obscured fraction that is constant with bolometric luminosity (Fig. \ref{fig:overview}, left panel); (b) recent observations of the COSMOS field have revealed candidate radio-loud AGN at $z\approx7-8$ that are obscured in the UV \citep[][]{Endsley22,Endsley23, Lambrides23}; 
the very large bolometric luminosities of these sources ($L_\mathrm{bol}\approx10^{47}\,\ergs$) together with the small FoV of the observations, implies an AGN obscured fraction that is consistent to the one inferred for bright LRDs. 

The large abundance of LRDs/obscured AGN has implications for their host halo masses. If obscuration were solely a viewing angle effect \citep[e.g.,][]{Antonucci93}, then we would expect LRDs to reside in the same environments as (equally bolometrically bright) UV-luminous quasars. Two decades of quasar clustering studies have constrained the masses of the dark matter halos hosting UV-luminous quasars at $0\lesssim z\lesssim6$ to be in the range $M_h\approx10^{12}-10^{13.5}\,\msun$ \citep[e.g.,][]{porciani_2004,croom2005,porciani_norberg2006,shen2007,shen2009,ross2013, eftekharzadeh2015, arita2023, eilers2024}, with little to no dependence on quasar luminosity \citep[e.g.,][]{adelberger_steidel2005, porciani_norberg2006, shen2009}. 
Several models have been put forward to interpret this host mass range in physical terms \citep[e.g.,][]{hopkins2007,fanidakis2013,caplar2015}{}{}. Whatever the reason for these characteristic host masses, it is striking that the number density of available host halos at these masses drops very quickly below the measured abundance of LRDs as redshift increases. At $z\approx6$, for example, LRDs (with $L_\mathrm{bol}>10^{45.5}\,\ergs$) are $\approx5\times$ more abundant than $10^{12}\,\msun$ halos (Fig. \ref{fig:mocks}, top right panel) and can occupy all halos above the threshold mass of $\Tilde{M}_\mathrm{h}>10^{11.7}\,\msun$. This implies that at these redshifts the host masses of LRDs are likely lower than the ones of UV-luminous quasars, even when matching them in $L_\mathrm{bol}$ space.

In Fig. \ref{fig:qhmfs}, we have presented a quantitative analysis of this argument at the two redshifts for which we have constraints on the clustering of bright ($L_\mathrm{bol}\approx10^{47}\,\ergs$), high-$z$ unobscured quasars: $z=4$ \citep[][]{shen2007} and $z=6.25$ \citep[][]{eilers2024}. We used the model developed in \citet{pizzati2024a, pizzati2024b} to measure the UV-luminous quasar host mass functions (QHMFs) at these two redshifts. While these QHMFs are well-constrained by clustering measurements only for the bright quasar population, we can extend them to also include the contribution of faint ($L_\mathrm{bol}\gtrsim10^{45.5}\,\ergs$) quasars by using the empirical quasar luminosity-halo mass relations obtained by \citet{pizzati2024a,pizzati2024b}. These relations are fit to the faint end of the quasar luminosity function, and hence they correctly reproduce the demographic properties of the faint quasar population. While we find minor differences in the luminosity dependence of the QHMFs at the two redshifts considered, we reach a general fundamental conclusion that is valid for faint and bright sources alike: \textit{the dark matter halos hosting UV-luminous quasars at $z\gtrsim4$ are too rare to accommodate the large number density of LRDs}.

What are the implications of these findings? 
If LRDs live in more common and hence less biased halos than those of unobscured quasars, then they may represent an intrinsically different population of accreting SMBHs arising in the early Universe. This population may be tracing a distinct phase in the co-evolutionary sequence of SMBHs and galaxies, similarly to what has been argued for type-2/reddened quasars at low redshifts \citep[e.g.,][]{allevato2014, cordova_rosado2024}.
In this scenario, the scaling relations between, e.g., black hole and halo/galaxy host masses need to be intrinsically different for LRDs and standard unobscured quasars, because similar black hole masses correspond to very different halo (and hence galaxy) masses. In particular, LRDs likely host SMBHs that are overmassive with respect to the coeval black hole-halo/galaxy mass scaling relations for unobscured quasars. 
Another possibility that has been put forward by several independent works to explain the enigmatic features of LRD SEDs \citep[e.g.,][]{greene2024, pacucci_narayan2024, lambrides2024} is that LRDs are accreting at rates that are larger than the critical Eddington limit. In this latter case, LRDs could represent the early stages of black hole accretion and growth that are predicted by many theoretical models of SMBH evolution \citep[e.g.,][]{trinca2023,li2024_bh_growth, lupi2024}. Interestingly, this would have direct implications for the clustering of LRDs, because a low duty cycle (that is necessary for super-Eddington accretion) would only be possible if LRDs lived in very low mass halos ($M_\mathrm{h}\approx10^{11}\,\msun$ at $z\approx6$; Sec. \ref{sec:lrd_mocks}).

Alternatively, these results may be telling us that key properties of LRDs, such as their bolometric luminosities and the relative contribution of the central AGN and the host galaxy to their observed SEDs, have yet to be properly characterized.
Indeed, the assumption on which our discussion is based, is that LRDs have the same bolometric luminosities as high-$z$ UV-luminous quasars ($L_\mathrm{bol}\approx10^{45}-10^{47}\,\ergs$). Currently, the bolometric luminosities of LRDs are estimated by their (dereddened) broad emission lines or by fitting AGN templates to their SEDs. In both cases, the resulting $L_\mathrm{bol}$ hinge on the assumption that the rest-frame optical continuum is dominated by AGN light \citep[see e.g.,][]{akins2024}. 
If the contribution of the host galaxy to the rest-frame optical continuum emission (and possibly broad lines; see, e.g., \citealt{baggen2024}) is non-negligible, then the inferred black hole masses and bolometric luminosities could change significantly. Several puzzling features of LRDs, such as their X-ray weakness \citep[][]{ananna2024, yue2024, maiolino2024} 
and (possibly) the lack of a hot dust torus \citep[][]{wang2024, perez-gonzalez2024, akins2024, iani2024} and UV variability \citep[][]{mitsuru2024}, 
point to the fact 
that LRD bolometric luminosities could be vastly overestimated. The presence of an evolved stellar population dominating (part of) the rest-frame optical is also suggested by the detection of a Balmer break in some LRD spectra \citep[e.g.,][but see \citealt{inayoshi_maiolino2024}]{wang2024, kokorev2024b}, 
although the large densities and stellar masses required to match the observed LRD luminosities remain a significant challenge to a purely stellar interpretation of LRD SEDs \citep[e.g.,][but see \citealt{baggen2024}]{greene2024, akins2024}. 
Regardless of the exact AGN contribution to these SEDs, if LRDs are not as bright as standard, UV-luminous quasars then they would naturally reside in lower mass halos, and they could easily be accommodated in the large number of $z\gtrsim6$ host halos with masses of $M_\mathrm{h}\approx10^{11}-10^{11.5}\,\msun$. 

In this work, we have primarily focused on the population of LRDs whose inferred SMBH masses and bolometric luminosities largely overlap with those of UV-luminous quasars. However, JWST has uncovered a much larger population of AGN with broad optical (H$\alpha$ or H$\beta$) lines, which are not necessarily reddened at optical wavelengths and hence do not respect the LRD selection criteria. Interestingly, the abundance of these broad-line AGN are even larger than the ones of LRDs: \citet{maiolino2024}, \citet{harikane2023}, and \citet{taylor2024} find the number densities for these sources to be in the range $10^{-3}-10^{-5}\,\cMpc^{-3}\mathrm{mag}^{-1}$ ($4\lesssim z \lesssim 7$). The intrinsic bolometric luminosities and SMBH masses of these broad-line AGN (that are not reddened in the rest-frame optical) are not as extreme as the ones of LRDs/reddened AGN \citep[e.g.,][]{harikane2023, taylor2024}. However, these sources can still reach UV magnitudes of $M_\mathrm{UV}\approx-22$ and bolometric luminosities of $L_\mathrm{bol}\approx10^{45.5}\,\ergs$, which are close to the ones of the faintest UV-selected quasars known at $z\gtrsim4$ \citep[][]{matsuoka2022}. Given their number densities, these broad-line AGN overshoot the extrapolation of the UV-selected quasar luminosity functions by factors that are comparable to (or even higher than) those found for LRDs (Sec. \ref{sec:lrd_abundance}). Hence, similar arguments to the ones presented in our analysis apply to this larger AGN population: their abundance suggests that they live in halos that are likely less massive than those of comparably luminous UV-selected quasars, implying that they obey fundamentally different scaling relations. While a proper comparison between UV-selected quasars and JWST AGN is only possible for the LRD population with large inferred bolometric luminosities and SMBH masses, it is interesting to investigate the host mass distributions, duty cycles, and scaling relations of this larger population of faint broad-line AGN.

Ultimately, a measurement of the clustering of LRDs and other broad-line AGN will constrain such properties and test the conclusions that we have drawn in this work. Recent arguments on the clustering of these objects rely on single detections of AGN in close proximity \citep[][]{lin_aspire2024, tanaka2025}, on spectroscopic detections of galaxies in a single LRD field \citep[][]{schindler2025}, and on cross-correlating photometrically-selected galaxies and LRDs \citep[][]{arita2025}. In this work (Fig. \ref{fig:mocks}), we have shown that a convincing measurement of LRD clustering can be made by using JWST NIRCam/WFSS observations of several LRD fields to extract a cross-correlation function between LRDs and \OIII line-emitting galaxies (see also \citealt{matthee2025} for recent results based on a similar approach). We have suggested that, by putting together $\approx10$ different fields, it is possible to infer the characteristic host halo mass of LRDs with an accuracy of $\log_{10} M_\mathrm{h}\approx0.1-0.3$. 
In order to perform this kind of measurement, one would need to observe several fields containing LRDs using a NIRCam grism filter covering the \OIII doublet. Interestingly, such observations already exist for a fraction of the broad-line AGN in the sample of \citet{matthee2024}: JWST surveys such as CONGRESS (GO3577) and GTO4540/GTO4549 are performing NIRCam/WFSS observations of the GOODS-N and GOODS-S fields, which contains $\approx10$ broad-line AGN from the \citet{matthee2024} sample. So a first step towards determining the clustering of these enigmatic sources at $z\gtrsim5$ is already feasible with current data. Future JWST programs will be able to deliver the same kind of observations for samples of LRDs with quasar-like inferred bolometric luminosities and SMBH masses. By comparing the host halo masses resulting from these measurements to the different scenarios discussed in Sec. \ref{sec:lrd_mocks}, it will be possible to get fundamental insights into the properties of these objects.

At the same time, the clustering of the faint, UV-luminous quasar population at high redshifts is also largely unconstrained. By using the same strategy and targeting faint quasar fields with NIRCam/WFSS, it will also be possible to determine their clustering and host masses. This would test our model predictions (Fig. \ref{fig:qhmfs}) and determine the luminosity dependence of quasar clustering at high-$z$, effectively constraining the scaling relation between the quasar bolometric luminosity and the host halo mass. Even more importantly, it would create a benchmark to which the LRD population can be effectively compared, allowing us to investigate the nature of quasar activity and SMBH populations in the early Universe.

\section*{Acknowledgements}

We acknowledge helpful conversations with Junya Arita, Jorryt Matthee, Jacob Shen,  Fengwu Sun, Marta Volonteri, Minghao Yue, and Ben Wang. We are grateful to the FLAMINGO-10k team for making their simulation available.  
We are also grateful to the ENIGMA group at
UC Santa Barbara and Leiden University for discussion of an early version of this manuscript. 
JFH and EP acknowledge support from the European Research Council (ERC) under the European
Union’s Horizon 2020 research and innovation program (grant agreement No 885301).
JTS is supported by the Deutsche Forschungsgemeinschaft (DFG, German Research Foundation) - Project number 518006966.
FW acknowledges support from NSF Grant AST-2308258.
This work is partly supported by funding from the European Union’s Horizon 2020 research and innovation programme under the Marie Skłodowska-Curie grant agreement No 860744 (BiD4BESt). 
This work used the DiRAC Memory Intensive service (Cosma8) at the University of Durham, which is part of the STFC DiRAC HPC Facility (\url{www.dirac.ac.uk}). Access to DiRAC resources was granted through a Director’s Discretionary Time allocation in 2023/24, under the auspices of the UKRI-funded DiRAC Federation Project. The equipment was funded by BEIS capital funding via
STFC capital grants ST/K00042X/1, ST/P002293/1, ST/R002371/1 and ST/S002502/1,
Durham University and STFC operations grant ST/R000832/1. DiRAC is part of the
National e-Infrastructure.

\section*{Data Availability}
The derived data generated in this research will be shared on reasonable requests to the corresponding author.



\bibliographystyle{mnras}
\bibliography{biblio} 





\bsp	
\label{lastpage}
\end{document}

%% file: additional_tex/pack.tex
\usepackage{newtxtext,newtxmath}

\usepackage[T1]{fontenc}
\usepackage{ae,aecompl}
\usepackage[utf8]{inputenc}

\usepackage{graphicx}	

\usepackage{amsmath}	
\usepackage{amssymb}	
\usepackage{comment}
\usepackage{siunitx}
\usepackage{booktabs}
\usepackage{gensymb}
\usepackage{threeparttable} 

\usepackage{xcolor}

%% file: additional_tex/definitions.tex
\renewcommand{\d}{\mathrm{d}}

\let\oldnabla\nabla
\renewcommand{\nabla}{\vec{\oldnabla}}

\def\be{\begin{equation}}
\def\ee{\end{equation}}
\newcommand\code[1]{\textsc{\MakeLowercase{#1}}}

\def\gsim{\lower.5ex\hbox{\gtsima}} 
\def\lsim{\lower.5ex\hbox{\ltsima}} 
\def\gtsima{$\; \buildrel > \over \sim \;$} 
\def\ltsima{$\; \buildrel < \over \sim \;$} \def\gsim{\lower.5ex\hbox{\gtsima}} 
\def\lsim{\lower.5ex\hbox{\ltsima}} 
\def\simgt{\lower.5ex\hbox{\gtsima}} 
\def\simlt{\lower.5ex\hbox{\ltsima}}

\def\msun{{\rm M}_{\odot}}

\def\kms{\,\rm km\,s^{-1}}

\def\ergs{{\rm erg}\,{\rm s}^{-1}}

\def\S*{$\Sigma_{\rm SFR}$}

\def\OIII{\hbox{[O~$\scriptstyle\rm III $]~}}

\def\cMpc{{\rm cMpc}}
\def\cGpc{{\rm cGpc}}

%% file: final_version.bbl
\begin{thebibliography}{}
\makeatletter
\relax
\def\mn@urlcharsother{\let\do\@makeother \do\$\do\&\do\#\do\^\do\_\do\%\do\~}
\def\mn@doi{\begingroup\mn@urlcharsother \@ifnextchar [ {\mn@doi@} {\mn@doi@[]}}
\def\mn@doi@[#1]#2{\def\@tempa{#1}\ifx\@tempa\@empty \href {http://dx.doi.org/#2} {doi:#2}\else \href {http://dx.doi.org/#2} {#1}\fi \endgroup}
\def\mn@eprint#1#2{\mn@eprint@#1:#2::\@nil}
\def\mn@eprint@arXiv#1{\href {http://arxiv.org/abs/#1} {{\tt arXiv:#1}}}
\def\mn@eprint@dblp#1{\href {http://dblp.uni-trier.de/rec/bibtex/#1.xml} {dblp:#1}}
\def\mn@eprint@#1:#2:#3:#4\@nil{\def\@tempa {#1}\def\@tempb {#2}\def\@tempc {#3}\ifx \@tempc \@empty \let \@tempc \@tempb \let \@tempb \@tempa \fi \ifx \@tempb \@empty \def\@tempb {arXiv}\fi \@ifundefined {mn@eprint@\@tempb}{\@tempb:\@tempc}{\expandafter \expandafter \csname mn@eprint@\@tempb\endcsname \expandafter{\@tempc}}}

\bibitem[\protect\citeauthoryear{{Abbott} et~al.,}{{Abbott} et~al.}{2022}]{abbott_des2022}
{Abbott} T.~M.~C.,  et~al., 2022, \mn@doi [\prd] {10.1103/PhysRevD.105.023520}, \href {https://ui.adsabs.harvard.edu/abs/2022PhRvD.105b3520A} {105, 023520}

\bibitem[\protect\citeauthoryear{{Adelberger} \& {Steidel}}{{Adelberger} \& {Steidel}}{2005}]{adelberger_steidel2005}
{Adelberger} K.~L.,  {Steidel} C.~C.,  2005, \mn@doi [\apj] {10.1086/431789}, \href {https://ui.adsabs.harvard.edu/abs/2005ApJ...630...50A} {630, 50}

\bibitem[\protect\citeauthoryear{{Aird} et~al.,}{{Aird} et~al.}{2015}]{aird2015}
{Aird} J.,  et~al., 2015, \mn@doi [\apj] {10.1088/0004-637X/815/1/66}, \href {https://ui.adsabs.harvard.edu/abs/2015ApJ...815...66A} {815, 66}

\bibitem[\protect\citeauthoryear{{Akins} et~al.,}{{Akins} et~al.}{2024}]{akins2024}
{Akins} H.~B.,  et~al., 2024, \mn@doi [arXiv e-prints] {10.48550/arXiv.2406.10341}, \href {https://ui.adsabs.harvard.edu/abs/2024arXiv240610341A} {p. arXiv:2406.10341}

\bibitem[\protect\citeauthoryear{{Akiyama} et~al.,}{{Akiyama} et~al.}{2018}]{akiyama2018}
{Akiyama} M.,  et~al., 2018, \mn@doi [\pasj] {10.1093/pasj/psx091}, \href {https://ui.adsabs.harvard.edu/abs/2018PASJ...70S..34A} {70, S34}

\bibitem[\protect\citeauthoryear{{Allevato} et~al.,}{{Allevato} et~al.}{2014}]{allevato2014}
{Allevato} V.,  et~al., 2014, \mn@doi [\apj] {10.1088/0004-637X/796/1/4}, \href {https://ui.adsabs.harvard.edu/abs/2014ApJ...796....4A} {796, 4}

\bibitem[\protect\citeauthoryear{{Ananna}, {Bogd{\'a}n}, {Kov{\'a}cs}, {Natarajan}  \& {Hickox}}{{Ananna} et~al.}{2024}]{ananna2024}
{Ananna} T.~T.,  {Bogd{\'a}n} {\'A}.,  {Kov{\'a}cs} O.~E.,  {Natarajan} P.,   {Hickox} R.~C.,  2024, \mn@doi [\apjl] {10.3847/2041-8213/ad5669}, \href {https://ui.adsabs.harvard.edu/abs/2024ApJ...969L..18A} {969, L18}

\bibitem[\protect\citeauthoryear{{Antonucci}}{{Antonucci}}{1993}]{Antonucci93}
{Antonucci} R.,  1993, \mn@doi [\araa] {10.1146/annurev.aa.31.090193.002353}, \href {http://adsabs.harvard.edu/abs/1993ARA\%26A..31..473A} {31, 473}

\bibitem[\protect\citeauthoryear{{Arita} et~al.,}{{Arita} et~al.}{2023}]{arita2023}
{Arita} J.,  et~al., 2023, \mn@doi [\apj] {10.3847/1538-4357/ace43a}, \href {https://ui.adsabs.harvard.edu/abs/2023ApJ...954..210A} {954, 210}

\bibitem[\protect\citeauthoryear{{Arita}, {Kashikawa}, {Onoue}, {Yoshioka}, {Takeda}, {Hoshi}  \& {Shimizu}}{{Arita} et~al.}{2025}]{arita2025}
{Arita} J.,  {Kashikawa} N.,  {Onoue} M.,  {Yoshioka} T.,  {Takeda} Y.,  {Hoshi} H.,   {Shimizu} S.,  2025, \mn@doi [\mnras] {10.1093/mnras/stae2765}, \href {https://ui.adsabs.harvard.edu/abs/2025MNRAS.536.3677A} {536, 3677}

\bibitem[\protect\citeauthoryear{{Baggen} et~al.,}{{Baggen} et~al.}{2024}]{baggen2024}
{Baggen} J. F.~W.,  et~al., 2024, \mn@doi [\apjl] {10.3847/2041-8213/ad90b8}, \href {https://ui.adsabs.harvard.edu/abs/2024ApJ...977L..13B} {977, L13}

\bibitem[\protect\citeauthoryear{{Bardeen}, {Bond}, {Kaiser}  \& {Szalay}}{{Bardeen} et~al.}{1986}]{bardeen1986}
{Bardeen} J.~M.,  {Bond} J.~R.,  {Kaiser} N.,   {Szalay} A.~S.,  1986, \mn@doi [\apj] {10.1086/164143}, \href {https://ui.adsabs.harvard.edu/abs/1986ApJ...304...15B} {304, 15}

\bibitem[\protect\citeauthoryear{{Behroozi}, {Wechsler}, {Hearin}  \& {Conroy}}{{Behroozi} et~al.}{2019}]{universe_machine}
{Behroozi} P.,  {Wechsler} R.~H.,  {Hearin} A.~P.,   {Conroy} C.,  2019, \mn@doi [\mnras] {10.1093/mnras/stz1182}, \href {https://ui.adsabs.harvard.edu/abs/2019MNRAS.488.3143B} {488, 3143}

\bibitem[\protect\citeauthoryear{{Bennett}, {Sijacki}, {Costa}, {Laporte}  \& {Witten}}{{Bennett} et~al.}{2024}]{Bennett2024}
{Bennett} J.~S.,  {Sijacki} D.,  {Costa} T.,  {Laporte} N.,   {Witten} C.,  2024, \mn@doi [\mnras] {10.1093/mnras/stad3179}, \href {https://ui.adsabs.harvard.edu/abs/2024MNRAS.527.1033B} {527, 1033}

\bibitem[\protect\citeauthoryear{{Bezanson} et~al.,}{{Bezanson} et~al.}{2024}]{bezanson2022}
{Bezanson} R.,  et~al., 2024, \mn@doi [\apj] {10.3847/1538-4357/ad66cf}, \href {https://ui.adsabs.harvard.edu/abs/2024ApJ...974...92B} {974, 92}

\bibitem[\protect\citeauthoryear{{Bogd{\'a}n} et~al.,}{{Bogd{\'a}n} et~al.}{2024}]{bogdan2024}
{Bogd{\'a}n} {\'A}.,  et~al., 2024, \mn@doi [Nature Astronomy] {10.1038/s41550-023-02111-9}, \href {https://ui.adsabs.harvard.edu/abs/2024NatAs...8..126B} {8, 126}

\bibitem[\protect\citeauthoryear{{Caplar}, {Lilly}  \& {Trakhtenbrot}}{{Caplar} et~al.}{2015}]{caplar2015}
{Caplar} N.,  {Lilly} S.~J.,   {Trakhtenbrot} B.,  2015, \mn@doi [\apj] {10.1088/0004-637X/811/2/148}, \href {https://ui.adsabs.harvard.edu/abs/2015ApJ...811..148C} {811, 148}

\bibitem[\protect\citeauthoryear{{Circosta} et~al.,}{{Circosta} et~al.}{2019}]{Circosta19}
{Circosta} C.,  et~al., 2019, \mn@doi [\aap] {10.1051/0004-6361/201834426}, \href {https://ui.adsabs.harvard.edu/abs/2019A&A...623A.172C} {623, A172}

\bibitem[\protect\citeauthoryear{{C{\'o}rdova Rosado} et~al.,}{{C{\'o}rdova Rosado} et~al.}{2024}]{cordova_rosado2024}
{C{\'o}rdova Rosado} R.,  et~al., 2024, \mn@doi [arXiv e-prints] {10.48550/arXiv.2409.08314}, \href {https://ui.adsabs.harvard.edu/abs/2024arXiv240908314C} {p. arXiv:2409.08314}

\bibitem[\protect\citeauthoryear{{Costa}}{{Costa}}{2024}]{costa2024}
{Costa} T.,  2024, \mn@doi [\mnras] {10.1093/mnras/stae1157}, \href {https://ui.adsabs.harvard.edu/abs/2024MNRAS.531..930C} {531, 930}

\bibitem[\protect\citeauthoryear{{Croom}, {Smith}, {Boyle}, {Shanks}, {Miller}, {Outram}  \& {Loaring}}{{Croom} et~al.}{2004}]{croom2004}
{Croom} S.~M.,  {Smith} R.~J.,  {Boyle} B.~J.,  {Shanks} T.,  {Miller} L.,  {Outram} P.~J.,   {Loaring} N.~S.,  2004, \mn@doi [\mnras] {10.1111/j.1365-2966.2004.07619.x}, \href {https://ui.adsabs.harvard.edu/abs/2004MNRAS.349.1397C} {349, 1397}

\bibitem[\protect\citeauthoryear{{Croom} et~al.,}{{Croom} et~al.}{2005}]{croom2005}
{Croom} S.~M.,  et~al., 2005, \mn@doi [\mnras] {10.1111/j.1365-2966.2004.08379.x}, \href {https://ui.adsabs.harvard.edu/abs/2005MNRAS.356..415C} {356, 415}

\bibitem[\protect\citeauthoryear{{D'Amato} et~al.,}{{D'Amato} et~al.}{2020}]{DAmato20}
{D'Amato} Q.,  et~al., 2020, \mn@doi [\aap] {10.1051/0004-6361/201936175}, \href {https://ui.adsabs.harvard.edu/abs/2020A&A...636A..37D} {636, A37}

\bibitem[\protect\citeauthoryear{{Davies} et~al.,}{{Davies} et~al.}{2018}]{davies2018}
{Davies} F.~B.,  et~al., 2018, \mn@doi [\apj] {10.3847/1538-4357/aad6dc}, \href {https://ui.adsabs.harvard.edu/abs/2018ApJ...864..142D} {864, 142}

\bibitem[\protect\citeauthoryear{{Davies}, {Hennawi}  \& {Eilers}}{{Davies} et~al.}{2019}]{Davies2019b}
{Davies} F.~B.,  {Hennawi} J.~F.,   {Eilers} A.-C.,  2019, \mn@doi [\apjl] {10.3847/2041-8213/ab42e3}, \href {https://ui.adsabs.harvard.edu/abs/2019ApJ...884L..19D} {884, L19}

\bibitem[\protect\citeauthoryear{{Davies}, {Hennawi}  \& {Eilers}}{{Davies} et~al.}{2020}]{davies2020}
{Davies} F.~B.,  {Hennawi} J.~F.,   {Eilers} A.-C.,  2020, \mn@doi [\mnras] {10.1093/mnras/stz3303}, \href {https://ui.adsabs.harvard.edu/abs/2020MNRAS.493.1330D} {493, 1330}

\bibitem[\protect\citeauthoryear{{Dayal} et~al.,}{{Dayal} et~al.}{2024}]{dayal2024}
{Dayal} P.,  et~al., 2024, \mn@doi [arXiv e-prints] {10.48550/arXiv.2401.11242}, \href {https://ui.adsabs.harvard.edu/abs/2024arXiv240111242D} {p. arXiv:2401.11242}

\bibitem[\protect\citeauthoryear{{Diemer}}{{Diemer}}{2018}]{colossus_diemer2018}
{Diemer} B.,  2018, \mn@doi [\apjs] {10.3847/1538-4365/aaee8c}, \href {https://ui.adsabs.harvard.edu/abs/2018ApJS..239...35D} {239, 35}

\bibitem[\protect\citeauthoryear{{Durodola}, {Pacucci}  \& {Hickox}}{{Durodola} et~al.}{2024}]{durodola2024}
{Durodola} E.,  {Pacucci} F.,   {Hickox} R.~C.,  2024, \mn@doi [arXiv e-prints] {10.48550/arXiv.2406.10329}, \href {https://ui.adsabs.harvard.edu/abs/2024arXiv240610329D} {p. arXiv:2406.10329}

\bibitem[\protect\citeauthoryear{{Eftekharzadeh} et~al.,}{{Eftekharzadeh} et~al.}{2015}]{eftekharzadeh2015}
{Eftekharzadeh} S.,  et~al., 2015, \mn@doi [\mnras] {10.1093/mnras/stv1763}, \href {https://ui.adsabs.harvard.edu/abs/2015MNRAS.453.2779E} {453, 2779}

\bibitem[\protect\citeauthoryear{{Eilers}, {Hennawi}  \& {Davies}}{{Eilers} et~al.}{2018}]{Eilers2018b}
{Eilers} A.-C.,  {Hennawi} J.~F.,   {Davies} F.~B.,  2018, \mn@doi [\apj] {10.3847/1538-4357/aae081}, \href {http://adsabs.harvard.edu/abs/2018ApJ...867...30E} {867, 30}

\bibitem[\protect\citeauthoryear{{Eilers} et~al.,}{{Eilers} et~al.}{2020}]{Eilers2020}
{Eilers} A.-C.,  et~al., 2020, \mn@doi [\apj] {10.3847/1538-4357/aba52e}, \href {https://ui.adsabs.harvard.edu/abs/2020ApJ...900...37E} {900, 37}

\bibitem[\protect\citeauthoryear{{Eilers} et~al.,}{{Eilers} et~al.}{2024}]{eilers2024}
{Eilers} A.-C.,  et~al., 2024, \mn@doi [\apj] {10.3847/1538-4357/ad778b}, \href {https://ui.adsabs.harvard.edu/abs/2024ApJ...974..275E} {974, 275}

\bibitem[\protect\citeauthoryear{{Endsley} et~al.,}{{Endsley} et~al.}{2022}]{Endsley22}
{Endsley} R.,  et~al., 2022, \mn@doi [\mnras] {10.1093/mnras/stac737}, \href {https://ui.adsabs.harvard.edu/abs/2022MNRAS.512.4248E} {512, 4248}

\bibitem[\protect\citeauthoryear{{Endsley} et~al.,}{{Endsley} et~al.}{2023}]{Endsley23}
{Endsley} R.,  et~al., 2023, \mn@doi [\mnras] {10.1093/mnras/stad266}, \href {https://ui.adsabs.harvard.edu/abs/2023MNRAS.520.4609E} {520, 4609}

\bibitem[\protect\citeauthoryear{{Fan}, {Ba{\~n}ados}  \& {Simcoe}}{{Fan} et~al.}{2023}]{fan2022}
{Fan} X.,  {Ba{\~n}ados} E.,   {Simcoe} R.~A.,  2023, \mn@doi [\araa] {10.1146/annurev-astro-052920-102455}, \href {https://ui.adsabs.harvard.edu/abs/2023ARA&A..61..373F} {61, 373}

\bibitem[\protect\citeauthoryear{{Fanidakis}, {Macci{\`o}}, {Baugh}, {Lacey}  \& {Frenk}}{{Fanidakis} et~al.}{2013}]{fanidakis2013}
{Fanidakis} N.,  {Macci{\`o}} A.~V.,  {Baugh} C.~M.,  {Lacey} C.~G.,   {Frenk} C.~S.,  2013, \mn@doi [\mnras] {10.1093/mnras/stt1567}, \href {https://ui.adsabs.harvard.edu/abs/2013MNRAS.436..315F} {436, 315}

\bibitem[\protect\citeauthoryear{{Furtak} et~al.,}{{Furtak} et~al.}{2024}]{furtak2024}
{Furtak} L.~J.,  et~al., 2024, \mn@doi [\nat] {10.1038/s41586-024-07184-8}, \href {https://ui.adsabs.harvard.edu/abs/2024Natur.628...57F} {628, 57}

\bibitem[\protect\citeauthoryear{{Garc{\'\i}a-Vergara}, {Hennawi}, {Barrientos}  \& {Arrigoni Battaia}}{{Garc{\'\i}a-Vergara} et~al.}{2019}]{garcia-vergara2019}
{Garc{\'\i}a-Vergara} C.,  {Hennawi} J.~F.,  {Barrientos} L.~F.,   {Arrigoni Battaia} F.,  2019, \mn@doi [\apj] {10.3847/1538-4357/ab4d52}, \href {https://ui.adsabs.harvard.edu/abs/2019ApJ...886...79G} {886, 79}

\bibitem[\protect\citeauthoryear{{Gehrels}}{{Gehrels}}{1986}]{gehrels86}
{Gehrels} N.,  1986, \mn@doi [\apj] {10.1086/164079}, \href {https://ui.adsabs.harvard.edu/abs/1986ApJ...303..336G} {303, 336}

\bibitem[\protect\citeauthoryear{{Giallongo} et~al.,}{{Giallongo} et~al.}{2019}]{giallongo2019}
{Giallongo} E.,  et~al., 2019, \mn@doi [\apj] {10.3847/1538-4357/ab39e1}, \href {https://ui.adsabs.harvard.edu/abs/2019ApJ...884...19G} {884, 19}

\bibitem[\protect\citeauthoryear{{Gilli} et~al.,}{{Gilli} et~al.}{2022}]{Gilli2022}
{Gilli} R.,  et~al., 2022, \mn@doi [\aap] {10.1051/0004-6361/202243708}, \href {https://ui.adsabs.harvard.edu/abs/2022A&A...666A..17G} {666, A17}

\bibitem[\protect\citeauthoryear{{Glikman} et~al.,}{{Glikman} et~al.}{2018}]{glikman2018}
{Glikman} E.,  et~al., 2018, \mn@doi [\apj] {10.3847/1538-4357/aac5d8}, \href {https://ui.adsabs.harvard.edu/abs/2018ApJ...861...37G} {861, 37}

\bibitem[\protect\citeauthoryear{{Greene} et~al.,}{{Greene} et~al.}{2024}]{greene2024}
{Greene} J.~E.,  et~al., 2024, \mn@doi [\apj] {10.3847/1538-4357/ad1e5f}, \href {https://ui.adsabs.harvard.edu/abs/2024ApJ...964...39G} {964, 39}

\bibitem[\protect\citeauthoryear{{Haiman} \& {Hui}}{{Haiman} \& {Hui}}{2001}]{haiman_hui2001}
{Haiman} Z.,  {Hui} L.,  2001, \mn@doi [\apj] {10.1086/318330}, \href {https://ui.adsabs.harvard.edu/abs/2001ApJ...547...27H} {547, 27}

\bibitem[\protect\citeauthoryear{{Han}, {Jing}, {Wang}  \& {Wang}}{{Han} et~al.}{2012}]{hbt}
{Han} J.,  {Jing} Y.~P.,  {Wang} H.,   {Wang} W.,  2012, \mn@doi [\mnras] {10.1111/j.1365-2966.2012.22111.x}, \href {https://ui.adsabs.harvard.edu/abs/2012MNRAS.427.2437H} {427, 2437}

\bibitem[\protect\citeauthoryear{{Han}, {Cole}, {Frenk}, {Benitez-Llambay}  \& {Helly}}{{Han} et~al.}{2018}]{hbt_plus}
{Han} J.,  {Cole} S.,  {Frenk} C.~S.,  {Benitez-Llambay} A.,   {Helly} J.,  2018, \mn@doi [\mnras] {10.1093/mnras/stx2792}, \href {https://ui.adsabs.harvard.edu/abs/2018MNRAS.474..604H} {474, 604}

\bibitem[\protect\citeauthoryear{{Harikane} et~al.,}{{Harikane} et~al.}{2023}]{harikane2023}
{Harikane} Y.,  et~al., 2023, \mn@doi [\apj] {10.3847/1538-4357/ad029e}, \href {https://ui.adsabs.harvard.edu/abs/2023ApJ...959...39H} {959, 39}

\bibitem[\protect\citeauthoryear{{He} et~al.,}{{He} et~al.}{2018}]{he2018}
{He} W.,  et~al., 2018, \mn@doi [\pasj] {10.1093/pasj/psx129}, \href {https://ui.adsabs.harvard.edu/abs/2018PASJ...70S..33H} {70, S33}

\bibitem[\protect\citeauthoryear{{Hickox} et~al.,}{{Hickox} et~al.}{2011}]{hickox2011}
{Hickox} R.~C.,  et~al., 2011, \mn@doi [\apj] {10.1088/0004-637X/731/2/117}, \href {https://ui.adsabs.harvard.edu/abs/2011ApJ...731..117H} {731, 117}

\bibitem[\protect\citeauthoryear{{Hopkins}, {Hernquist}, {Martini}, {Cox}, {Robertson}, {Di Matteo}  \& {Springel}}{{Hopkins} et~al.}{2005}]{Hopkins05}
{Hopkins} P.~F.,  {Hernquist} L.,  {Martini} P.,  {Cox} T.~J.,  {Robertson} B.,  {Di Matteo} T.,   {Springel} V.,  2005, \mn@doi [The Astrophysical Journal] {10.1086/431146}, \href {https://ui.adsabs.harvard.edu/abs/2005ApJ...625L..71H} {625, L71}

\bibitem[\protect\citeauthoryear{{Hopkins}, {Lidz}, {Hernquist}, {Coil}, {Myers}, {Cox}  \& {Spergel}}{{Hopkins} et~al.}{2007}]{hopkins2007}
{Hopkins} P.~F.,  {Lidz} A.,  {Hernquist} L.,  {Coil} A.~L.,  {Myers} A.~D.,  {Cox} T.~J.,   {Spergel} D.~N.,  2007, \mn@doi [\apj] {10.1086/517512}, \href {https://ui.adsabs.harvard.edu/abs/2007ApJ...662..110H} {662, 110}

\bibitem[\protect\citeauthoryear{{Iani} et~al.,}{{Iani} et~al.}{2024}]{iani2024}
{Iani} E.,  et~al., 2024, \mn@doi [arXiv e-prints] {10.48550/arXiv.2406.18207}, \href {https://ui.adsabs.harvard.edu/abs/2024arXiv240618207I} {p. arXiv:2406.18207}

\bibitem[\protect\citeauthoryear{{Inayoshi} \& {Ichikawa}}{{Inayoshi} \& {Ichikawa}}{2024}]{inayoshi2024}
{Inayoshi} K.,  {Ichikawa} K.,  2024, \mn@doi [\apjl] {10.3847/2041-8213/ad74e2}, \href {https://ui.adsabs.harvard.edu/abs/2024ApJ...973L..49I} {973, L49}

\bibitem[\protect\citeauthoryear{{Inayoshi} \& {Maiolino}}{{Inayoshi} \& {Maiolino}}{2025}]{inayoshi_maiolino2024}
{Inayoshi} K.,  {Maiolino} R.,  2025, \mn@doi [\apjl] {10.3847/2041-8213/adaebd}, \href {https://ui.adsabs.harvard.edu/abs/2025ApJ...980L..27I} {980, L27}

\bibitem[\protect\citeauthoryear{{Inayoshi}, {Onoue}, {Sugahara}, {Inoue}  \& {Ho}}{{Inayoshi} et~al.}{2022}]{inayoshi2022}
{Inayoshi} K.,  {Onoue} M.,  {Sugahara} Y.,  {Inoue} A.~K.,   {Ho} L.~C.,  2022, \mn@doi [\apjl] {10.3847/2041-8213/ac6f01}, \href {https://ui.adsabs.harvard.edu/abs/2022ApJ...931L..25I} {931, L25}

\bibitem[\protect\citeauthoryear{{Kaiser}}{{Kaiser}}{1984}]{kaiser1984}
{Kaiser} N.,  1984, \mn@doi [\apjl] {10.1086/184341}, \href {https://ui.adsabs.harvard.edu/abs/1984ApJ...284L...9K} {284, L9}

\bibitem[\protect\citeauthoryear{{Kashino}, {Lilly}, {Matthee}, {Eilers}, {Mackenzie}, {Bordoloi}  \& {Simcoe}}{{Kashino} et~al.}{2023}]{kashino2022}
{Kashino} D.,  {Lilly} S.~J.,  {Matthee} J.,  {Eilers} A.-C.,  {Mackenzie} R.,  {Bordoloi} R.,   {Simcoe} R.~A.,  2023, \mn@doi [\apj] {10.3847/1538-4357/acc588}, \href {https://ui.adsabs.harvard.edu/abs/2023ApJ...950...66K} {950, 66}

\bibitem[\protect\citeauthoryear{{Khrykin}, {Hennawi}, {McQuinn}  \& {Worseck}}{{Khrykin} et~al.}{2016}]{khrykin2016}
{Khrykin} I.~S.,  {Hennawi} J.~F.,  {McQuinn} M.,   {Worseck} G.,  2016, \mn@doi [\apj] {10.3847/0004-637X/824/2/133}, \href {https://ui.adsabs.harvard.edu/abs/2016ApJ...824..133K} {824, 133}

\bibitem[\protect\citeauthoryear{{Khrykin}, {Hennawi}  \& {Worseck}}{{Khrykin} et~al.}{2019a}]{khrykin2019}
{Khrykin} I.~S.,  {Hennawi} J.~F.,   {Worseck} G.,  2019a, \mn@doi [\mnras] {10.1093/mnras/stz135}, \href {https://ui.adsabs.harvard.edu/abs/2019MNRAS.484.3897K} {484, 3897}

\bibitem[\protect\citeauthoryear{{Khrykin}, {Hennawi}  \& {Worseck}}{{Khrykin} et~al.}{2019b}]{khrykin2019b}
{Khrykin} I.~S.,  {Hennawi} J.~F.,   {Worseck} G.,  2019b, \mn@doi [\mnras] {10.1093/mnras/stz135}, \href {https://ui.adsabs.harvard.edu/abs/2019MNRAS.484.3897K} {484, 3897}

\bibitem[\protect\citeauthoryear{{Killi} et~al.,}{{Killi} et~al.}{2024}]{killi2024}
{Killi} M.,  et~al., 2024, \mn@doi [\aap] {10.1051/0004-6361/202348857}, \href {https://ui.adsabs.harvard.edu/abs/2024A&A...691A..52K} {691, A52}

\bibitem[\protect\citeauthoryear{{Kocevski} et~al.,}{{Kocevski} et~al.}{2023}]{kocevski2023}
{Kocevski} D.~D.,  et~al., 2023, \mn@doi [\apjl] {10.3847/2041-8213/ace5a0}, \href {https://ui.adsabs.harvard.edu/abs/2023ApJ...954L...4K} {954, L4}

\bibitem[\protect\citeauthoryear{{Kocevski} et~al.,}{{Kocevski} et~al.}{2024}]{kocevski2024}
{Kocevski} D.~D.,  et~al., 2024, \mn@doi [arXiv e-prints] {10.48550/arXiv.2404.03576}, \href {https://ui.adsabs.harvard.edu/abs/2024arXiv240403576K} {p. arXiv:2404.03576}

\bibitem[\protect\citeauthoryear{{Kokorev} et~al.,}{{Kokorev} et~al.}{2023}]{kokorev2023}
{Kokorev} V.,  et~al., 2023, \mn@doi [\apjl] {10.3847/2041-8213/ad037a}, \href {https://ui.adsabs.harvard.edu/abs/2023ApJ...957L...7K} {957, L7}

\bibitem[\protect\citeauthoryear{{Kokorev} et~al.,}{{Kokorev} et~al.}{2024a}]{kokorev2024}
{Kokorev} V.,  et~al., 2024a, \mn@doi [\apj] {10.3847/1538-4357/ad4265}, \href {https://ui.adsabs.harvard.edu/abs/2024ApJ...968...38K} {968, 38}

\bibitem[\protect\citeauthoryear{{Kokorev} et~al.,}{{Kokorev} et~al.}{2024b}]{kokorev2024b}
{Kokorev} V.,  et~al., 2024b, \mn@doi [\apj] {10.3847/1538-4357/ad7d03}, \href {https://ui.adsabs.harvard.edu/abs/2024ApJ...975..178K} {975, 178}

\bibitem[\protect\citeauthoryear{{Kokubo} \& {Harikane}}{{Kokubo} \& {Harikane}}{2024}]{mitsuru2024}
{Kokubo} M.,  {Harikane} Y.,  2024, \mn@doi [arXiv e-prints] {10.48550/arXiv.2407.04777}, \href {https://ui.adsabs.harvard.edu/abs/2024arXiv240704777K} {p. arXiv:2407.04777}

\bibitem[\protect\citeauthoryear{{Kugel} et~al.,}{{Kugel} et~al.}{2023}]{FlamingoII}
{Kugel} R.,  et~al., 2023, \mn@doi [\mnras] {10.1093/mnras/stad2540}, \href {https://ui.adsabs.harvard.edu/abs/2023MNRAS.526.6103K} {526, 6103}

\bibitem[\protect\citeauthoryear{{Kulkarni}, {Worseck}  \& {Hennawi}}{{Kulkarni} et~al.}{2019}]{kulkarni2019}
{Kulkarni} G.,  {Worseck} G.,   {Hennawi} J.~F.,  2019, \mn@doi [\mnras] {10.1093/mnras/stz1493}, \href {https://ui.adsabs.harvard.edu/abs/2019MNRAS.488.1035K} {488, 1035}

\bibitem[\protect\citeauthoryear{{Labbe} et~al.,}{{Labbe} et~al.}{2025}]{labbe2023}
{Labbe} I.,  et~al., 2025, \mn@doi [\apj] {10.3847/1538-4357/ad3551}, \href {https://ui.adsabs.harvard.edu/abs/2025ApJ...978...92L} {978, 92}

\bibitem[\protect\citeauthoryear{{Lacy}, {Ridgway}, {Sajina}, {Petric}, {Gates}, {Urrutia}  \& {Storrie-Lombardi}}{{Lacy} et~al.}{2015}]{lacy2015}
{Lacy} M.,  {Ridgway} S.~E.,  {Sajina} A.,  {Petric} A.~O.,  {Gates} E.~L.,  {Urrutia} T.,   {Storrie-Lombardi} L.~J.,  2015, \mn@doi [\apj] {10.1088/0004-637X/802/2/102}, \href {https://ui.adsabs.harvard.edu/abs/2015ApJ...802..102L} {802, 102}

\bibitem[\protect\citeauthoryear{{Lambrides} et~al.,}{{Lambrides} et~al.}{2024a}]{lambrides2024}
{Lambrides} E.,  et~al., 2024a, arXiv e-prints, \href {https://ui.adsabs.harvard.edu/abs/2024arXiv240913047L} {p. arXiv:2409.13047}

\bibitem[\protect\citeauthoryear{{Lambrides} et~al.,}{{Lambrides} et~al.}{2024b}]{Lambrides23}
{Lambrides} E.,  et~al., 2024b, \mn@doi [\apjl] {10.3847/2041-8213/ad11ee}, \href {https://ui.adsabs.harvard.edu/abs/2024ApJ...961L..25L} {961, L25}

\bibitem[\protect\citeauthoryear{{Li} et~al.,}{{Li} et~al.}{2024}]{li2024_bh_growth}
{Li} W.,  et~al., 2024, \mn@doi [\apj] {10.3847/1538-4357/ad46f9}, \href {https://ui.adsabs.harvard.edu/abs/2024ApJ...969...69L} {969, 69}

\bibitem[\protect\citeauthoryear{{Li}, {Inayoshi}, {Chen}, {Ichikawa}  \& {Ho}}{{Li} et~al.}{2025}]{li2024}
{Li} Z.,  {Inayoshi} K.,  {Chen} K.,  {Ichikawa} K.,   {Ho} L.~C.,  2025, \mn@doi [\apj] {10.3847/1538-4357/ada5fb}, \href {https://ui.adsabs.harvard.edu/abs/2025ApJ...980...36L} {980, 36}

\bibitem[\protect\citeauthoryear{{Lidz}, {Hopkins}, {Cox}, {Hernquist}  \& {Robertson}}{{Lidz} et~al.}{2006}]{lidz2006}
{Lidz} A.,  {Hopkins} P.~F.,  {Cox} T.~J.,  {Hernquist} L.,   {Robertson} B.,  2006, \mn@doi [\apj] {10.1086/500444}, \href {https://ui.adsabs.harvard.edu/abs/2006ApJ...641...41L} {641, 41}

\bibitem[\protect\citeauthoryear{{Lin} et~al.,}{{Lin} et~al.}{2024}]{lin_aspire2024}
{Lin} X.,  et~al., 2024, \mn@doi [\apj] {10.3847/1538-4357/ad6565}, \href {https://ui.adsabs.harvard.edu/abs/2024ApJ...974..147L} {974, 147}

\bibitem[\protect\citeauthoryear{{Lupi}, {Trinca}, {Volonteri}, {Dotti}  \& {Mazzucchelli}}{{Lupi} et~al.}{2024}]{lupi2024}
{Lupi} A.,  {Trinca} A.,  {Volonteri} M.,  {Dotti} M.,   {Mazzucchelli} C.,  2024, \mn@doi [\aap] {10.1051/0004-6361/202451249}, \href {https://ui.adsabs.harvard.edu/abs/2024A&A...689A.128L} {689, A128}

\bibitem[\protect\citeauthoryear{{Lynden-Bell}}{{Lynden-Bell}}{1969}]{lynden_bell1969}
{Lynden-Bell} D.,  1969, \mn@doi [\nat] {10.1038/223690a0}, \href {https://ui.adsabs.harvard.edu/abs/1969Natur.223..690L} {223, 690}

\bibitem[\protect\citeauthoryear{{Madau}, {Giallongo}, {Grazian}  \& {Haardt}}{{Madau} et~al.}{2024}]{madau2024}
{Madau} P.,  {Giallongo} E.,  {Grazian} A.,   {Haardt} F.,  2024, \mn@doi [\apj] {10.3847/1538-4357/ad5ce8}, \href {https://ui.adsabs.harvard.edu/abs/2024ApJ...971...75M} {971, 75}

\bibitem[\protect\citeauthoryear{{Maiolino} et~al.,}{{Maiolino} et~al.}{2024a}]{maiolino2024}
{Maiolino} R.,  et~al., 2024a, \mn@doi [\nat] {10.1038/s41586-024-07052-5}, \href {https://ui.adsabs.harvard.edu/abs/2024Natur.627...59M} {627, 59}

\bibitem[\protect\citeauthoryear{{Maiolino} et~al.,}{{Maiolino} et~al.}{2024b}]{Maiolino23}
{Maiolino} R.,  et~al., 2024b, \mn@doi [\aap] {10.1051/0004-6361/202347640}, \href {https://ui.adsabs.harvard.edu/abs/2024A&A...691A.145M} {691, A145}

\bibitem[\protect\citeauthoryear{{Martini} \& {Weinberg}}{{Martini} \& {Weinberg}}{2001}]{martini2001}
{Martini} P.,  {Weinberg} D.~H.,  2001, \mn@doi [\apj] {10.1086/318331}, \href {https://ui.adsabs.harvard.edu/abs/2001ApJ...547...12M} {547, 12}

\bibitem[\protect\citeauthoryear{{Matsuoka} et~al.,}{{Matsuoka} et~al.}{2018}]{matsuoka2018}
{Matsuoka} Y.,  et~al., 2018, \mn@doi [\apj] {10.3847/1538-4357/aaee7a}, \href {https://ui.adsabs.harvard.edu/abs/2018ApJ...869..150M} {869, 150}

\bibitem[\protect\citeauthoryear{{Matsuoka} et~al.,}{{Matsuoka} et~al.}{2022}]{matsuoka2022}
{Matsuoka} Y.,  et~al., 2022, \mn@doi [\apjs] {10.3847/1538-4365/ac3d31}, \href {https://ui.adsabs.harvard.edu/abs/2022ApJS..259...18M} {259, 18}

\bibitem[\protect\citeauthoryear{{Matsuoka} et~al.,}{{Matsuoka} et~al.}{2023}]{matsuoka2023}
{Matsuoka} Y.,  et~al., 2023, \mn@doi [\apjl] {10.3847/2041-8213/acd69f}, \href {https://ui.adsabs.harvard.edu/abs/2023ApJ...949L..42M} {949, L42}

\bibitem[\protect\citeauthoryear{{Matthee}, {Mackenzie}, {Simcoe}, {Kashino}, {Lilly}, {Bordoloi}  \& {Eilers}}{{Matthee} et~al.}{2023}]{matthee2023}
{Matthee} J.,  {Mackenzie} R.,  {Simcoe} R.~A.,  {Kashino} D.,  {Lilly} S.~J.,  {Bordoloi} R.,   {Eilers} A.-C.,  2023, \mn@doi [\apj] {10.3847/1538-4357/acc846}, \href {https://ui.adsabs.harvard.edu/abs/2023ApJ...950...67M} {950, 67}

\bibitem[\protect\citeauthoryear{{Matthee} et~al.,}{{Matthee} et~al.}{2024a}]{matthee2025}
{Matthee} J.,  et~al., 2024a, \mn@doi [arXiv e-prints] {10.48550/arXiv.2412.02846}, \href {https://ui.adsabs.harvard.edu/abs/2024arXiv241202846M} {p. arXiv:2412.02846}

\bibitem[\protect\citeauthoryear{{Matthee} et~al.,}{{Matthee} et~al.}{2024b}]{matthee2024}
{Matthee} J.,  et~al., 2024b, \mn@doi [\apj] {10.3847/1538-4357/ad2345}, \href {https://ui.adsabs.harvard.edu/abs/2024ApJ...963..129M} {963, 129}

\bibitem[\protect\citeauthoryear{{Mazzolari} et~al.,}{{Mazzolari} et~al.}{2024}]{mazzolari2024}
{Mazzolari} G.,  et~al., 2024, \mn@doi [arXiv e-prints] {10.48550/arXiv.2408.15615}, \href {https://ui.adsabs.harvard.edu/abs/2024arXiv240815615M} {p. arXiv:2408.15615}

\bibitem[\protect\citeauthoryear{{McGreer}, {Fan}, {Jiang}  \& {Cai}}{{McGreer} et~al.}{2018}]{mcgreer2018}
{McGreer} I.~D.,  {Fan} X.,  {Jiang} L.,   {Cai} Z.,  2018, \mn@doi [\aj] {10.3847/1538-3881/aaaab4}, \href {https://ui.adsabs.harvard.edu/abs/2018AJ....155..131M} {155, 131}

\bibitem[\protect\citeauthoryear{{Merloni} et~al.,}{{Merloni} et~al.}{2014}]{Merloni14}
{Merloni} A.,  et~al., 2014, \mn@doi [Monthly Notices of the Royal Astronomical Society] {10.1093/mnras/stt2149}, \href {https://ui.adsabs.harvard.edu/abs/2014MNRAS.437.3550M} {437, 3550}

\bibitem[\protect\citeauthoryear{{Mo} \& {White}}{{Mo} \& {White}}{1996}]{mo_white1996}
{Mo} H.~J.,  {White} S.~D.~M.,  1996, \mn@doi [\mnras] {10.1093/mnras/282.2.347}, \href {https://ui.adsabs.harvard.edu/abs/1996MNRAS.282..347M} {282, 347}

\bibitem[\protect\citeauthoryear{{Mu{\~n}oz}, {Mirocha}, {Furlanetto}  \& {Sabti}}{{Mu{\~n}oz} et~al.}{2023}]{munoz2023}
{Mu{\~n}oz} J.~B.,  {Mirocha} J.,  {Furlanetto} S.,   {Sabti} N.,  2023, \mn@doi [\mnras] {10.1093/mnrasl/slad115}, \href {https://ui.adsabs.harvard.edu/abs/2023MNRAS.526L..47M} {526, L47}

\bibitem[\protect\citeauthoryear{{Myers} et~al.,}{{Myers} et~al.}{2006}]{myers2006}
{Myers} A.~D.,  et~al., 2006, \mn@doi [\apj] {10.1086/499093}, \href {https://ui.adsabs.harvard.edu/abs/2006ApJ...638..622M} {638, 622}

\bibitem[\protect\citeauthoryear{{Ni}, {Di Matteo}, {Gilli}, {Croft}, {Feng}  \& {Norman}}{{Ni} et~al.}{2020}]{Ni2020}
{Ni} Y.,  {Di Matteo} T.,  {Gilli} R.,  {Croft} R. A.~C.,  {Feng} Y.,   {Norman} C.,  2020, \mn@doi [\mnras] {10.1093/mnras/staa1313}, \href {https://ui.adsabs.harvard.edu/abs/2020MNRAS.495.2135N} {495, 2135}

\bibitem[\protect\citeauthoryear{{Niida} et~al.,}{{Niida} et~al.}{2020}]{niida2020}
{Niida} M.,  et~al., 2020, \mn@doi [\apj] {10.3847/1538-4357/abbe11}, \href {https://ui.adsabs.harvard.edu/abs/2020ApJ...904...89N} {904, 89}

\bibitem[\protect\citeauthoryear{{Onken}, {Wolf}, {Bian}, {Fan}, {Hon}, {Raithel}, {Tisserand}  \& {Lai}}{{Onken} et~al.}{2022}]{onken2021}
{Onken} C.~A.,  {Wolf} C.,  {Bian} F.,  {Fan} X.,  {Hon} W.~J.,  {Raithel} D.,  {Tisserand} P.,   {Lai} S.,  2022, \mn@doi [\mnras] {10.1093/mnras/stac051}, \href {https://ui.adsabs.harvard.edu/abs/2022MNRAS.511..572O} {511, 572}

\bibitem[\protect\citeauthoryear{{Pacucci} \& {Loeb}}{{Pacucci} \& {Loeb}}{2022}]{pacucci2022}
{Pacucci} F.,  {Loeb} A.,  2022, in American Astronomical Society Meeting \#240. p. 213.04

\bibitem[\protect\citeauthoryear{{Pacucci} \& {Narayan}}{{Pacucci} \& {Narayan}}{2024}]{pacucci_narayan2024}
{Pacucci} F.,  {Narayan} R.,  2024, \mn@doi [arXiv e-prints] {10.48550/arXiv.2407.15915}, \href {https://ui.adsabs.harvard.edu/abs/2024arXiv240715915P} {p. arXiv:2407.15915}

\bibitem[\protect\citeauthoryear{{Pacucci}, {Nguyen}, {Carniani}, {Maiolino}  \& {Fan}}{{Pacucci} et~al.}{2023}]{pacucci2023_mbhmstar}
{Pacucci} F.,  {Nguyen} B.,  {Carniani} S.,  {Maiolino} R.,   {Fan} X.,  2023, \mn@doi [\apjl] {10.3847/2041-8213/ad0158}, \href {https://ui.adsabs.harvard.edu/abs/2023ApJ...957L...3P} {957, L3}

\bibitem[\protect\citeauthoryear{{Padovani} et~al.,}{{Padovani} et~al.}{2017}]{padovani2017}
{Padovani} P.,  et~al., 2017, \mn@doi [\aapr] {10.1007/s00159-017-0102-9}, \href {https://ui.adsabs.harvard.edu/abs/2017A&ARv..25....2P} {25, 2}

\bibitem[\protect\citeauthoryear{{Pan}, {Jiang}, {Fan}, {Wu}  \& {Yang}}{{Pan} et~al.}{2022}]{pan2022}
{Pan} Z.,  {Jiang} L.,  {Fan} X.,  {Wu} J.,   {Yang} J.,  2022, \mn@doi [\apj] {10.3847/1538-4357/ac5aab}, \href {https://ui.adsabs.harvard.edu/abs/2022ApJ...928..172P} {928, 172}

\bibitem[\protect\citeauthoryear{{P{\'e}rez-Gonz{\'a}lez} et~al.,}{{P{\'e}rez-Gonz{\'a}lez} et~al.}{2024}]{perez-gonzalez2024}
{P{\'e}rez-Gonz{\'a}lez} P.~G.,  et~al., 2024, \mn@doi [\apj] {10.3847/1538-4357/ad38bb}, \href {https://ui.adsabs.harvard.edu/abs/2024ApJ...968....4P} {968, 4}

\bibitem[\protect\citeauthoryear{{Petter}, {Hickox}, {Alexander}, {Myers}, {Geach}, {Whalen}  \& {Andonie}}{{Petter} et~al.}{2023}]{petter2023}
{Petter} G.~C.,  {Hickox} R.~C.,  {Alexander} D.~M.,  {Myers} A.~D.,  {Geach} J.~E.,  {Whalen} K.~E.,   {Andonie} C.~P.,  2023, \mn@doi [\apj] {10.3847/1538-4357/acb7ef}, \href {https://ui.adsabs.harvard.edu/abs/2023ApJ...946...27P} {946, 27}

\bibitem[\protect\citeauthoryear{{Pizzati}, {Hennawi}, {Schaye}  \& {Schaller}}{{Pizzati} et~al.}{2024a}]{pizzati2024a}
{Pizzati} E.,  {Hennawi} J.~F.,  {Schaye} J.,   {Schaller} M.,  2024a, \mn@doi [\mnras] {10.1093/mnras/stae329}, \href {https://ui.adsabs.harvard.edu/abs/2024MNRAS.528.4466P} {528, 4466}

\bibitem[\protect\citeauthoryear{{Pizzati} et~al.,}{{Pizzati} et~al.}{2024b}]{pizzati2024b}
{Pizzati} E.,  et~al., 2024b, \mn@doi [\mnras] {10.1093/mnras/stae2307}, \href {https://ui.adsabs.harvard.edu/abs/2024MNRAS.534.3155P} {534, 3155}

\bibitem[\protect\citeauthoryear{{Porciani} \& {Norberg}}{{Porciani} \& {Norberg}}{2006}]{porciani_norberg2006}
{Porciani} C.,  {Norberg} P.,  2006, \mn@doi [\mnras] {10.1111/j.1365-2966.2006.10813.x}, \href {https://ui.adsabs.harvard.edu/abs/2006MNRAS.371.1824P} {371, 1824}

\bibitem[\protect\citeauthoryear{{Porciani}, {Magliocchetti}  \& {Norberg}}{{Porciani} et~al.}{2004}]{porciani_2004}
{Porciani} C.,  {Magliocchetti} M.,   {Norberg} P.,  2004, \mn@doi [\mnras] {10.1111/j.1365-2966.2004.08408.x}, \href {https://ui.adsabs.harvard.edu/abs/2004MNRAS.355.1010P} {355, 1010}

\bibitem[\protect\citeauthoryear{{Ren}, {Trenti}  \& {Di Matteo}}{{Ren} et~al.}{2020}]{ren_trenti2020}
{Ren} K.,  {Trenti} M.,   {Di Matteo} T.,  2020, \mn@doi [\apj] {10.3847/1538-4357/ab86ab}, \href {https://ui.adsabs.harvard.edu/abs/2020ApJ...894..124R} {894, 124}

\bibitem[\protect\citeauthoryear{{Richards} et~al.,}{{Richards} et~al.}{2006}]{richards2006}
{Richards} G.~T.,  et~al., 2006, \mn@doi [\aj] {10.1086/503559}, \href {https://ui.adsabs.harvard.edu/abs/2006AJ....131.2766R} {131, 2766}

\bibitem[\protect\citeauthoryear{{Ross} et~al.,}{{Ross} et~al.}{2009}]{ross2009}
{Ross} N.~P.,  et~al., 2009, \mn@doi [\apj] {10.1088/0004-637X/697/2/1634}, \href {https://ui.adsabs.harvard.edu/abs/2009ApJ...697.1634R} {697, 1634}

\bibitem[\protect\citeauthoryear{{Ross} et~al.,}{{Ross} et~al.}{2013}]{ross2013}
{Ross} N.~P.,  et~al., 2013, \mn@doi [\apj] {10.1088/0004-637X/773/1/14}, \href {https://ui.adsabs.harvard.edu/abs/2013ApJ...773...14R} {773, 14}

\bibitem[\protect\citeauthoryear{{Runnoe}, {Brotherton}  \& {Shang}}{{Runnoe} et~al.}{2012a}]{runnoe2012}
{Runnoe} J.~C.,  {Brotherton} M.~S.,   {Shang} Z.,  2012a, \mn@doi [\mnras] {10.1111/j.1365-2966.2012.20620.x}, \href {https://ui.adsabs.harvard.edu/abs/2012MNRAS.422..478R} {422, 478}

\bibitem[\protect\citeauthoryear{{Runnoe}, {Brotherton}  \& {Shang}}{{Runnoe} et~al.}{2012b}]{runnoe2012b}
{Runnoe} J.~C.,  {Brotherton} M.~S.,   {Shang} Z.,  2012b, \mn@doi [\mnras] {10.1111/j.1365-2966.2012.21644.x}, \href {https://ui.adsabs.harvard.edu/abs/2012MNRAS.426.2677R} {426, 2677}

\bibitem[\protect\citeauthoryear{{Salpeter}}{{Salpeter}}{1964}]{salpeter1964}
{Salpeter} E.~E.,  1964, \mn@doi [\apj] {10.1086/147973}, \href {https://ui.adsabs.harvard.edu/abs/1964ApJ...140..796S} {140, 796}

\bibitem[\protect\citeauthoryear{{Sanders}, {Soifer}, {Elias}, {Madore}, {Matthews}, {Neugebauer}  \& {Scoville}}{{Sanders} et~al.}{1988}]{Sanders88}
{Sanders} D.~B.,  {Soifer} B.~T.,  {Elias} J.~H.,  {Madore} B.~F.,  {Matthews} K.,  {Neugebauer} G.,   {Scoville} N.~Z.,  1988, \mn@doi [\apj] {10.1086/165983}, \href {http://adsabs.harvard.edu/abs/1988ApJ...325...74S} {325, 74}

\bibitem[\protect\citeauthoryear{{Sanders}, {Phinney}, {Neugebauer}, {Soifer}  \& {Matthews}}{{Sanders} et~al.}{1989}]{sanders1989}
{Sanders} D.~B.,  {Phinney} E.~S.,  {Neugebauer} G.,  {Soifer} B.~T.,   {Matthews} K.,  1989, \mn@doi [\apj] {10.1086/168094}, \href {https://ui.adsabs.harvard.edu/abs/1989ApJ...347...29S} {347, 29}

\bibitem[\protect\citeauthoryear{{Schaye} et~al.,}{{Schaye} et~al.}{2023}]{flamingo}
{Schaye} J.,  et~al., 2023, \mn@doi [\mnras] {10.1093/mnras/stad2419}, \href {https://ui.adsabs.harvard.edu/abs/2023MNRAS.526.4978S} {526, 4978}

\bibitem[\protect\citeauthoryear{{Schindler} et~al.,}{{Schindler} et~al.}{2019}]{schindler2019}
{Schindler} J.-T.,  et~al., 2019, \mn@doi [\apj] {10.3847/1538-4357/aaf86c}, \href {https://ui.adsabs.harvard.edu/abs/2019ApJ...871..258S} {871, 258}

\bibitem[\protect\citeauthoryear{{Schindler} et~al.,}{{Schindler} et~al.}{2023}]{schindler2023}
{Schindler} J.-T.,  et~al., 2023, \mn@doi [\apj] {10.3847/1538-4357/aca7ca}, \href {https://ui.adsabs.harvard.edu/abs/2023ApJ...943...67S} {943, 67}

\bibitem[\protect\citeauthoryear{{Schindler} et~al.,}{{Schindler} et~al.}{2024}]{schindler2025}
{Schindler} J.-T.,  et~al., 2024, \mn@doi [arXiv e-prints] {10.48550/arXiv.2411.11534}, \href {https://ui.adsabs.harvard.edu/abs/2024arXiv241111534S} {p. arXiv:2411.11534}

\bibitem[\protect\citeauthoryear{{Scholtz} et~al.,}{{Scholtz} et~al.}{2023}]{scholtz2023}
{Scholtz} J.,  et~al., 2023, \mn@doi [arXiv e-prints] {10.48550/arXiv.2311.18731}, \href {https://ui.adsabs.harvard.edu/abs/2023arXiv231118731S} {p. arXiv:2311.18731}

\bibitem[\protect\citeauthoryear{{Shankar}, {Weinberg}  \& {Shen}}{{Shankar} et~al.}{2010a}]{shankar2010_lowz}
{Shankar} F.,  {Weinberg} D.~H.,   {Shen} Y.,  2010a, \mn@doi [\mnras] {10.1111/j.1365-2966.2010.16801.x}, \href {https://ui.adsabs.harvard.edu/abs/2010MNRAS.406.1959S} {406, 1959}

\bibitem[\protect\citeauthoryear{{Shankar}, {Crocce}, {Miralda-Escud{\'e}}, {Fosalba}  \& {Weinberg}}{{Shankar} et~al.}{2010b}]{shankar_2010}
{Shankar} F.,  {Crocce} M.,  {Miralda-Escud{\'e}} J.,  {Fosalba} P.,   {Weinberg} D.~H.,  2010b, \mn@doi [\apj] {10.1088/0004-637X/718/1/231}, \href {https://ui.adsabs.harvard.edu/abs/2010ApJ...718..231S} {718, 231}

\bibitem[\protect\citeauthoryear{{Shen} et~al.,}{{Shen} et~al.}{2007}]{shen2007}
{Shen} Y.,  et~al., 2007, \mn@doi [\aj] {10.1086/513517}, \href {https://ui.adsabs.harvard.edu/abs/2007AJ....133.2222S} {133, 2222}

\bibitem[\protect\citeauthoryear{{Shen} et~al.,}{{Shen} et~al.}{2009}]{shen2009}
{Shen} Y.,  et~al., 2009, \mn@doi [\apj] {10.1088/0004-637X/697/2/1656}, \href {https://ui.adsabs.harvard.edu/abs/2009ApJ...697.1656S} {697, 1656}

\bibitem[\protect\citeauthoryear{{Shen}, {Hopkins}, {Faucher-Gigu{\`e}re}, {Alexander}, {Richards}, {Ross}  \& {Hickox}}{{Shen} et~al.}{2020}]{shen2020}
{Shen} X.,  {Hopkins} P.~F.,  {Faucher-Gigu{\`e}re} C.-A.,  {Alexander} D.~M.,  {Richards} G.~T.,  {Ross} N.~P.,   {Hickox} R.~C.,  2020, \mn@doi [\mnras] {10.1093/mnras/staa1381}, \href {https://ui.adsabs.harvard.edu/abs/2020MNRAS.495.3252S} {495, 3252}

\bibitem[\protect\citeauthoryear{{Soltan}}{{Soltan}}{1982}]{Soltan1982}
{Soltan} A.,  1982, \mn@doi [\mnras] {10.1093/mnras/200.1.115}, \href {https://ui.adsabs.harvard.edu/abs/1982MNRAS.200..115S} {200, 115}

\bibitem[\protect\citeauthoryear{{Tanaka} et~al.,}{{Tanaka} et~al.}{2024}]{tanaka2025}
{Tanaka} T.~S.,  et~al., 2024, \mn@doi [arXiv e-prints] {10.48550/arXiv.2412.14246}, \href {https://ui.adsabs.harvard.edu/abs/2024arXiv241214246T} {p. arXiv:2412.14246}

\bibitem[\protect\citeauthoryear{{Taylor} et~al.,}{{Taylor} et~al.}{2024}]{taylor2024}
{Taylor} A.~J.,  et~al., 2024, \mn@doi [arXiv e-prints] {10.48550/arXiv.2409.06772}, \href {https://ui.adsabs.harvard.edu/abs/2024arXiv240906772T} {p. arXiv:2409.06772}

\bibitem[\protect\citeauthoryear{{Timlin} et~al.,}{{Timlin} et~al.}{2018}]{timlin2018}
{Timlin} J.~D.,  et~al., 2018, \mn@doi [\apj] {10.3847/1538-4357/aab9ac}, \href {https://ui.adsabs.harvard.edu/abs/2018ApJ...859...20T} {859, 20}

\bibitem[\protect\citeauthoryear{{Tinker}, {Kravtsov}, {Klypin}, {Abazajian}, {Warren}, {Yepes}, {Gottl{\"o}ber}  \& {Holz}}{{Tinker} et~al.}{2008}]{tinker2008}
{Tinker} J.,  {Kravtsov} A.~V.,  {Klypin} A.,  {Abazajian} K.,  {Warren} M.,  {Yepes} G.,  {Gottl{\"o}ber} S.,   {Holz} D.~E.,  2008, \mn@doi [\apj] {10.1086/591439}, \href {https://ui.adsabs.harvard.edu/abs/2008ApJ...688..709T} {688, 709}

\bibitem[\protect\citeauthoryear{{Trinca}, {Schneider}, {Maiolino}, {Valiante}, {Graziani}  \& {Volonteri}}{{Trinca} et~al.}{2023}]{trinca2023}
{Trinca} A.,  {Schneider} R.,  {Maiolino} R.,  {Valiante} R.,  {Graziani} L.,   {Volonteri} M.,  2023, \mn@doi [\mnras] {10.1093/mnras/stac3768}, \href {https://ui.adsabs.harvard.edu/abs/2023MNRAS.519.4753T} {519, 4753}

\bibitem[\protect\citeauthoryear{{{\"U}bler} et~al.,}{{{\"U}bler} et~al.}{2023}]{ubler2023}
{{\"U}bler} H.,  et~al., 2023, \mn@doi [\aap] {10.1051/0004-6361/202346137}, \href {https://ui.adsabs.harvard.edu/abs/2023A&A...677A.145U} {677, A145}

\bibitem[\protect\citeauthoryear{{Ueda}, {Akiyama}, {Ohta}  \& {Miyaji}}{{Ueda} et~al.}{2003}]{Ueda03}
{Ueda} Y.,  {Akiyama} M.,  {Ohta} K.,   {Miyaji} T.,  2003, \mn@doi [\apj] {10.1086/378940}, \href {https://ui.adsabs.harvard.edu/abs/2003ApJ...598..886U} {598, 886}

\bibitem[\protect\citeauthoryear{{Ueda}, {Akiyama}, {Hasinger}, {Miyaji}  \& {Watson}}{{Ueda} et~al.}{2014}]{Ueda14}
{Ueda} Y.,  {Akiyama} M.,  {Hasinger} G.,  {Miyaji} T.,   {Watson} M.~G.,  2014, \mn@doi [\apj] {10.1088/0004-637X/786/2/104}, \href {https://ui.adsabs.harvard.edu/abs/2014ApJ...786..104U} {786, 104}

\bibitem[\protect\citeauthoryear{{Urry} \& {Padovani}}{{Urry} \& {Padovani}}{1995}]{UrryPadovani95}
{Urry} C.~M.,  {Padovani} P.,  1995, \mn@doi [\pasp] {10.1086/133630}, \href {http://adsabs.harvard.edu/abs/1995PASP..107..803U} {107, 803}

\bibitem[\protect\citeauthoryear{{Vito} et~al.,}{{Vito} et~al.}{2018}]{Vito18}
{Vito} F.,  et~al., 2018, \mn@doi [Monthly Notices of the Royal Astronomical Society] {10.1093/mnras/stx2486}, \href {https://ui.adsabs.harvard.edu/abs/2018MNRAS.473.2378V} {473, 2378}

\bibitem[\protect\citeauthoryear{{Vito}, {Di Mascia}, {Gallerani}, {Zana}, {Ferrara}, {Carniani}  \& {Gilli}}{{Vito} et~al.}{2022}]{Vito2022}
{Vito} F.,  {Di Mascia} F.,  {Gallerani} S.,  {Zana} T.,  {Ferrara} A.,  {Carniani} S.,   {Gilli} R.,  2022, \mn@doi [\mnras] {10.1093/mnras/stac1422}, \href {https://ui.adsabs.harvard.edu/abs/2022MNRAS.514.1672V} {514, 1672}

\bibitem[\protect\citeauthoryear{{Wang} et~al.,}{{Wang} et~al.}{2019}]{wang2019}
{Wang} F.,  et~al., 2019, \mn@doi [\apj] {10.3847/1538-4357/ab2be5}, \href {https://ui.adsabs.harvard.edu/abs/2019ApJ...884...30W} {884, 30}

\bibitem[\protect\citeauthoryear{{Wang} et~al.,}{{Wang} et~al.}{2023}]{aspire_wang2023}
{Wang} F.,  et~al., 2023, \mn@doi [\apjl] {10.3847/2041-8213/accd6f}, \href {https://ui.adsabs.harvard.edu/abs/2023ApJ...951L...4W} {951, L4}

\bibitem[\protect\citeauthoryear{{Wang} et~al.,}{{Wang} et~al.}{2024}]{wang2024}
{Wang} B.,  et~al., 2024, \mn@doi [arXiv e-prints] {10.48550/arXiv.2403.02304}, \href {https://ui.adsabs.harvard.edu/abs/2024arXiv240302304W} {p. arXiv:2403.02304}

\bibitem[\protect\citeauthoryear{{White}, {Martini}  \& {Cohn}}{{White} et~al.}{2008}]{white_2008}
{White} M.,  {Martini} P.,   {Cohn} J.~D.,  2008, \mn@doi [\mnras] {10.1111/j.1365-2966.2008.13817.x}, \href {https://ui.adsabs.harvard.edu/abs/2008MNRAS.390.1179W} {390, 1179}

\bibitem[\protect\citeauthoryear{{Worseck}, {Prochaska}, {Hennawi}  \& {McQuinn}}{{Worseck} et~al.}{2016}]{worseck2016}
{Worseck} G.,  {Prochaska} J.~X.,  {Hennawi} J.~F.,   {McQuinn} M.,  2016, \mn@doi [\apj] {10.3847/0004-637X/825/2/144}, \href {https://ui.adsabs.harvard.edu/abs/2016ApJ...825..144W} {825, 144}

\bibitem[\protect\citeauthoryear{{Worseck}, {Khrykin}, {Hennawi}, {Prochaska}  \& {Farina}}{{Worseck} et~al.}{2021}]{worseck2021}
{Worseck} G.,  {Khrykin} I.~S.,  {Hennawi} J.~F.,  {Prochaska} J.~X.,   {Farina} E.~P.,  2021, \mn@doi [\mnras] {10.1093/mnras/stab1685}, \href {https://ui.adsabs.harvard.edu/abs/2021MNRAS.505.5084W} {505, 5084}

\bibitem[\protect\citeauthoryear{{Wyithe} \& {Loeb}}{{Wyithe} \& {Loeb}}{2009}]{wyithe_loeb2009}
{Wyithe} J. S.~B.,  {Loeb} A.,  2009, \mn@doi [\mnras] {10.1111/j.1365-2966.2009.14647.x}, \href {https://ui.adsabs.harvard.edu/abs/2009MNRAS.395.1607W} {395, 1607}

\bibitem[\protect\citeauthoryear{{Yang}, {Mo}  \& {van den Bosch}}{{Yang} et~al.}{2003}]{yang2003}
{Yang} X.,  {Mo} H.~J.,   {van den Bosch} F.~C.,  2003, \mn@doi [\mnras] {10.1046/j.1365-8711.2003.06254.x}, \href {https://ui.adsabs.harvard.edu/abs/2003MNRAS.339.1057Y} {339, 1057}

\bibitem[\protect\citeauthoryear{{Yang} et~al.,}{{Yang} et~al.}{2016}]{yang2016}
{Yang} J.,  et~al., 2016, \mn@doi [\apj] {10.3847/0004-637X/829/1/33}, \href {https://ui.adsabs.harvard.edu/abs/2016ApJ...829...33Y} {829, 33}

\bibitem[\protect\citeauthoryear{{Yang} et~al.,}{{Yang} et~al.}{2023}]{yang2023_desi}
{Yang} J.,  et~al., 2023, \mn@doi [\apjs] {10.3847/1538-4365/acf99b}, \href {https://ui.adsabs.harvard.edu/abs/2023ApJS..269...27Y} {269, 27}

\bibitem[\protect\citeauthoryear{{York} et~al.,}{{York} et~al.}{2000}]{york2000}
{York} D.~G.,  et~al., 2000, \mn@doi [\aj] {10.1086/301513}, \href {https://ui.adsabs.harvard.edu/abs/2000AJ....120.1579Y} {120, 1579}

\bibitem[\protect\citeauthoryear{{Yu} \& {Tremaine}}{{Yu} \& {Tremaine}}{2002}]{yu_tremaine2002}
{Yu} Q.,  {Tremaine} S.,  2002, \mn@doi [\mnras] {10.1046/j.1365-8711.2002.05532.x}, \href {https://ui.adsabs.harvard.edu/abs/2002MNRAS.335..965Y} {335, 965}

\bibitem[\protect\citeauthoryear{{Yue} et~al.,}{{Yue} et~al.}{2024a}]{yue2024_eiger}
{Yue} M.,  et~al., 2024a, \mn@doi [\apj] {10.3847/1538-4357/ad3914}, \href {https://ui.adsabs.harvard.edu/abs/2024ApJ...966..176Y} {966, 176}

\bibitem[\protect\citeauthoryear{{Yue}, {Eilers}, {Ananna}, {Panagiotou}, {Kara}  \& {Miyaji}}{{Yue} et~al.}{2024b}]{yue2024}
{Yue} M.,  {Eilers} A.-C.,  {Ananna} T.~T.,  {Panagiotou} C.,  {Kara} E.,   {Miyaji} T.,  2024b, \mn@doi [\apjl] {10.3847/2041-8213/ad7eba}, \href {https://ui.adsabs.harvard.edu/abs/2024ApJ...974L..26Y} {974, L26}

\bibitem[\protect\citeauthoryear{{Zel'dovich} \& {Novikov}}{{Zel'dovich} \& {Novikov}}{1967}]{zeldovich_1964}
{Zel'dovich} Y.~B.,  {Novikov} I.~D.,  1967, \sovast, \href {https://ui.adsabs.harvard.edu/abs/1967SvA....10..602Z} {10, 602}

\bibitem[\protect\citeauthoryear{{{\v{D}}urov{\v{c}}{\'\i}kov{\'a}} et~al.,}{{{\v{D}}urov{\v{c}}{\'\i}kov{\'a}} et~al.}{2024}]{Durovcikova2024}
{{\v{D}}urov{\v{c}}{\'\i}kov{\'a}} D.,  et~al., 2024, \mn@doi [\apj] {10.3847/1538-4357/ad4888}, \href {https://ui.adsabs.harvard.edu/abs/2024ApJ...969..162D} {969, 162}

\makeatother
\end{thebibliography}
